\documentclass[aps,prd,preprint,tightenlines,preprintnumbers,showpacs,superscriptaddress,nofootinbib,eqsecnum,floatfix
]{revtex4-1}

\usepackage{bm}
\usepackage{graphicx}
\usepackage[pdftex]{color}
\usepackage{amsmath, amssymb}
\usepackage{hyperref}
\usepackage{dcolumn}
\usepackage{array}
\hypersetup{
    colorlinks=true,     
    linkcolor=blue,      
    citecolor=blue,      
    filecolor=blue,      
    urlcolor=blue        
}
\usepackage{rotating}
\usepackage{mathtools}
\usepackage{multirow}
\usepackage{slashed} 
\usepackage[normalem]{ulem} 

\definecolor{red}{rgb}{0.8,0.0,0.0}
\definecolor{green}{rgb}{0.0,0.6,0.0}
\definecolor{darkblue}{rgb}{0.0,0.1,0.7}
\definecolor{brown}{rgb}{0.6,0.1,0.0}
\definecolor{gray}{rgb}{0.6,0.6,0.6}
\definecolor{darkgreen}{rgb}{0.0, 0.545098, 0.0}
\definecolor{verydarkgreen}{rgb}{0.0, 0.4, 0.0}
\definecolor{veryverydarkgreen}{rgb}{0.0, 0.3, 0.0}
\definecolor{purple}{rgb}{0.5,0.0,0.5}
\definecolor{applegreen}{rgb}{0.55, 0.71, 0.0}
\definecolor{babypink} {rgb}{0.64, 0.44, 0.44}
\definecolor{orange}{rgb}{0.9,0.4,0.0}

\newcommand{\bi}{\begin{itemize}}
\newcommand{\ei}{\end{itemize}}

\newcommand{\be}{\begin{equation}}
\newcommand{\ee}{\end{equation}}

\newcommand{\bea}{\begin{eqnarray}}
\newcommand{\eea}{\end{eqnarray}}

\newcolumntype{P}[1]{>{\centering\arraybackslash}p{#1}} 

\def\Eq#1{Eq.~(\ref{#1})}

\newcommand{\spose}[1]{\hbox to 0pt{#1\hss}}
\newcommand{\inapprox}{\mathrel{\spose{\lower 3pt\hbox{$\mathchar"218$}} \raise 2.0pt\hbox{$\mathchar"232$}}}
\newcommand\qhat{\hat q}

\newcommand{\Dslash}{\ensuremath{D\kern-0.6em/\kern0.15em}} 
\newcommand{\tr}{\ensuremath{\mathop{\text{tr}}}} 
\def\svev#1{\left\langle #1\right\rangle}       

\newcommand{\GeV}{~\ensuremath{\text{GeV}}}   
\newcommand{\MeV}{~\ensuremath{\text{MeV}}}  

\newcommand{\PiLR}{\ensuremath{\Pi_{LR}}} 
\newcommand\CLR{\ensuremath{C_{LR}}}
\newcommand{\hCLR}{\ensuremath{\hat{C}_{LR}}}

\def\U1{{\rm U}(1)}

\newcommand{\4}{\ensuremath{\text{\textbf{4}}}} 
\newcommand{\6}{\ensuremath{\text{\textbf{6}}}} 

\def\ha{\hat{a}}

\def\hm{\hat{m}}

\begin{document}


\title{Radiative contribution to the composite-Higgs potential in a two-representation lattice model}



\author{Venkitesh Ayyar}
\affiliation{NERSC, Lawrence Berkeley National Lab  Berkeley, CA 94720 USA}


\author{Maarten Golterman}
\affiliation{Department of Physics and Astronomy, San Francisco State University, San Francisco,
CA 94132, USA}

\author{Daniel C.~Hackett}
\affiliation{Department of Physics, University of Colorado, Boulder, CO 80309, USA}

\author{William~I.~Jay}\email{wjay@fnal.gov}
\affiliation{Theoretical Physics Department, Fermi National Accelerator Laboratory, Batavia, Illinois, 60510, USA}

\author{Ethan T.~Neil}\email{ethan.neil@colorado.edu}
\affiliation{Department of Physics, University of Colorado, Boulder, CO 80309, USA}

\author{Yigal~Shamir}
\affiliation{Raymond and Beverly Sackler School of Physics and Astronomy,
Tel~Aviv University, 69978 Tel~Aviv, Israel}

\author{Benjamin Svetitsky}
\affiliation{Raymond and Beverly Sackler School of Physics and Astronomy,
Tel~Aviv University, 69978 Tel~Aviv, Israel}

\date{\today}

\begin{abstract}
Working in a two-representation lattice gauge theory that is close to a composite Higgs model, we calculate the low-energy constant $\CLR$ which controls the contribution of the electroweak gauge bosons to the Higgs potential.
In QCD, the corresponding low-energy constant governs the mass splitting of the pion multiplet.   Taking the continuum and chiral limits, we find that $\CLR$, in units of the pseudoscalar decay constant, is roughly of the same size as its QCD counterpart.
\end{abstract}

\pacs{
    11.15.Ha,   
    12.60.Rc
}
\preprint{FERMILAB-PUB-19-083-T}
\maketitle

\begin{flushleft}


\end{flushleft}


\section{Introduction \label{sec:intro}}
Composite Higgs theories of the ``Goldstone Higgs'' variety \cite{Georgi:1984af,Dugan:1984hq} use a weakly broken global symmetry to protect the Higgs from large mass renormalizations.
The models are often written down as effective field theories in the form of non-linear sigma models \cite{Contino:2010rs,Bellazzini:2014yua,Panico:2015jxa}.
The sigma model describes a set of exactly massless Nambu--Goldstone bosons that live in a coset manifold $G/H$.
This set contains the Higgs multiplet of the Standard Model (SM).
The Higgs potential then comes mainly from coupling to the electroweak
gauge bosons and to the top quark, via one-loop diagrams.
This potential should induce the Higgs phenomenon of the SM.

The sigma model is only an effective low-energy description with the correct symmetry
properties.
The coupling of the Higgs multiplet to the SM fields, which in turn yields the Higgs potential, is given by a number of low-energy constants.
These are, in principle, calculable if an ultraviolet completion of the theory is given.
We continue here a study of a model that is close to such an ultraviolet theory, one of a set catalogued by Ferretti and Karateev
\cite{Ferretti:2013kya,Ferretti:2014qta,Ferretti:2016upr,Cacciapaglia:2019bqz} that accommodate both a composite Higgs and a partially composite top quark~\cite{Kaplan:1991dc}.
For that theory, the Higgs potential was discussed in Refs.~\cite{Golterman:2015zwa,Ferretti:2016upr,Golterman:2017vdj}.

The model we study is an SU(4) gauge theory containing two multiplets of fermions.
The first consists of $N_f = 2$ Dirac fermions in the sextet representation of SU(4), which is the antisymmetric two-index representation---a real representation.
The second contains Dirac fermions in the fundamental representation of SU(4), again with $N_f = 2$.
A Goldstone multiplet arises when the global SU(4) symmetry carried by the sextet fermions is spontaneously broken to SO(4).
In Ferretti's composite Higgs model
\cite{Ferretti:2014qta},  5 flavors of Majorana fermions in the sextet representation of SU(4) give rise to a composite Higgs within an SU$(5)/$SO(5) coset.
Ferretti's model also
contains fundamental Dirac fermions, but with $N_f= 3$, to allow construction of a top partner.

When the fermions are weakly coupled to a gauge field outside this model, the Goldstone fields acquire a potential.
We present a lattice calculation of the low-energy constant $\CLR$ that enters this potential.
In Ferretti's model and in similar theories, the latter is given by
\cite{Contino:2010rs,Golterman:2014yha,Golterman:2015zwa}
\be
V_{\rm eff}(\Sigma) = C_{LR}\sum_Q\tr\left(Q\Sigma Q^*\Sigma^*\right)\ ,
\label{LEC}
\ee
where $\Sigma$ is the non-linear field representing
the multiplet of pseudo-Goldstone bosons.
The sum over $Q$ runs over the $SU(2)_L$ generators $gT^a_L$
and the hypercharge generator $g'Y$, with $g$ and $g'$ the electroweak
coupling constants of the SM\@.
In accordance with vacuum alignment \cite{Peskin:1980gc},
the low-energy constant $C_{LR}$ is positive
\cite{Witten:1983ut,Golterman:2014yha}, and the minimum of $V_{\rm eff}$ is attained at $\langle\Sigma\rangle={\bf 1}$.
The physics of \Eq{LEC} is analogous to the electromagnetic mass splitting between pions in QCD, a point which we revisit in the discussion.
Realization of the Higgs phenomenon requires $V_{\rm eff}$ to have
a negative curvature at $\Sigma={\bf 1}$.
This is expected to arise
from coupling to the top quark \cite{Agashe:2004rs,Ferretti:2014qta,Golterman:2015zwa,Ferretti:2016upr,Golterman:2017vdj}, which we do not treat here.
For phenomenological use of $\CLR$, see for example Ref.~\cite{DelDebbio:2017ini}.

We obtain $\CLR$ in terms of a correlation function of the ultraviolet theory~\cite{Das:1967it},
\begin{equation}
\CLR=16\pi^2\int\frac{d^4q}{(2\pi)^4}\,\PiLR(q_\mu).
\label{eq:CLR4dintegral}
\end{equation}
Here $\PiLR(q^2)$ is the transverse part of the current--current correlation function,
\be
\frac12\delta_{ab}\Pi_{\mu\nu}(q) = -\int {d^4 x} \,e^{iqx}\, \langle J_{\mu a}^L(x) J_{\nu b}^R(0)\rangle,
\label{PiLRdefn}
\ee
which defines $\PiLR(q^2)$ via the decomposition
\be
  \Pi_{\mu\nu}(q) =  (q^2 \delta_{\mu\nu} - q_\mu q_\nu)\Pi_{LR}(q^2) + q_\mu q_\nu\Pi'(q^2) .
  \label{PiLRdecomp}
\ee
The chiral currents, constructed from the sextet fermions, are
\bea
J^L_{\mu a}&=&\bar\psi\gamma_\mu(1-\gamma_5){T_a}\psi
= V_{\mu a}-A_{\mu a}\,,\nonumber\\[3pt]
J^R_{\mu a}&=&\bar\psi\gamma_\mu(1+\gamma_5){T_a}\psi
= V_{\mu a}+A_{\mu a}\,,
  \label{chiralCurrents}
\eea
where $T_a$ are the isospin generators.
In the chiral limit, where the low-energy constant in \Eq{LEC} is defined, the current correlator (\ref{PiLRdefn}) is automatically transverse.

In a previous exploratory study, we calculated $\CLR$ in the same theory but without the fundamental fermions \cite{DeGrand:2016htl}.
The present calculation differs from the earlier one in several ways:
\begin{enumerate}
\item
As noted, we have added fermions in the fundamental representation.
This brings the theory closer to
Ferretti's composite Higgs model \cite{Ferretti:2014qta}.
We do not actually simulate Ferretti's model for technical reasons, namely, the well-known difficulty of simulating lattice theories with anything other than an even number of Dirac flavors.
\item
Exact chirality is important when constructing the currents (\ref{chiralCurrents})
(see Sec.~\ref{sec:chiral}).
In Ref.~\cite{DeGrand:2016htl} we calculated the current correlators with valence overlap fermions.
Here we use a more economical prescription, constructing the correlators with valence staggered fermions.
The limited chiral symmetry of the latter is enough to guarantee the desired properties of the correlators.
(Other calculations of $\PiLR(q)$ in QCD and beyond \cite{Shintani:2008qe,Boyle:2009xi,Appelquist:2010xv} have used overlap or domain-wall fermions.
For calculations of the vector current two-point function
using staggered fermions, see Refs.~\cite{Aubin:2006xv,Chakraborty:2014mwa,Chakraborty:2016mwy,Borsanyi:2017zdw,Aubin:2018fog}.)

\item
Our previous work used only two ensembles, generated with different lattice actions (both based on improved Wilson fermions) but with roughly equal lattice spacings and similar physical properties.
While we took a chiral limit for the valence fermions, we did not attempt to take a continuum limit or to extrapolate to the chiral limit of the sea fermions.
In this work we measure $\CLR$ in ten ensembles with a range of sea masses and lattice spacings.
We have studied these ensembles at length \cite{Ayyar:2017qdf,Ayyar:2018zuk,Ayyar:2018glg}.
For each ensemble we have set a scale via the flow variable $t_0$ and measured the meson spectrum.
Hence we are now able to fit $\CLR$ as a function of lattice spacing and sea masses, and to take the continuum and chiral limits in the dynamical theory.
\end{enumerate}

Our paper is organized as follows.
In Sec.~\ref{sec:chiral} we review properties of the current correlation function $\PiLR$, the importance of chiral symmetry in its calculation, and its definition on the lattice with staggered fermions.
In Sec.~\ref{sec:numerical} we present the calculation of $\PiLR$ and its integral $\CLR$ on each ensemble, along with the extrapolation to massless valence fermions.
In Sec.~\ref{sec:limits} we take all the ensembles together in order to fit $\CLR$ and extrapolate it to the continuum and chiral limits.
We conclude with discussion of our results in Sec.~\ref{sec:discussion}.

\section{Chiral symmetry and staggered fermions \label{sec:chiral}}

\subsection{Chiral symmetry\label{sec:ch_cont}}

In infinite volume, the value of $\CLR$
depends {\em a priori\/} on the dynamical infrared scale $\Lambda$ of the theory,
on the fermion mass $m$, and on an ultraviolet cutoff $M$.
Chiral symmetry is essential in removing the effects of the ultraviolet cutoff from $\CLR$.
To see this, we recall \cite{DeGrand:2016htl} the operator product expansion
for the two-current correlator, which is, schematically,
\be
  \Pi_{\textit{XX}}(q^2;m) \sim 1+ \frac{m^2}{q^2}
     + \frac{g^2\svev{G_{\mu\nu}G_{\mu\nu}}+m\svev{\bar\psi\psi}}{q^4}
     + \frac{\Lambda^6}{q^6}+ \cdots\ ,
\label{OPEXX}
\ee
where $\textit{XX}=VV$ or $AA$.  Each term is to be multiplied
by a coefficient function that depends logarithmically on $q^2$.
In the difference $\Pi_{VV}-\Pi_{AA}$, the identity term drops out, as do all purely gluonic condensates, and we have
\be
  \PiLR(q^2;m) \sim \frac{m^2}{q^2} + \frac{m\Lambda^3}{q^4}
     + \frac{\Lambda^6}{q^6} + \cdots\ ,
\label{OPEPiLR}
\ee
where, for each power of $1/q^2$, we show only the leading dependence
on the fermion mass.

We introduce the ultraviolet cutoff $M$ into \Eq{eq:CLR4dintegral} via
\be
  \CLR(m;M) = \int_0^{M^2} dq^2\, q^2\, \PiLR(q^2;m) \ .
\label{Ccontcut}
\ee
Inserting the expansion (\ref{OPEPiLR}) we obtain
\be
  \CLR(m;M) \sim m^2M^2 + m\Lambda^3\log(M) + \Lambda^4 + O(1/M^2),
  \label{cutoffDep}
  \ee
where $\Lambda^4$ comes from the infrared part of the integral.
Hence $\CLR(m;M)$ is quadratically divergent for $m\not=0$.
Nonetheless, $\CLR(0;M)$ is finite in the $M\to\infty$ limit, giving the desired low-energy constant $\CLR$.
In other words, we must take the (valence) chiral limit
before the continuum limit.
This result could have been anticipated by noting that $\CLR$ is
an order parameter for the spontaneous breaking of chiral symmetry.

If the cutoff procedure breaks chiral symmetry, on the other hand, the constant term in \Eq{OPEXX} does not cancel between $VV$ and $AA$, and $\CLR(m\to0)$ remains (quartically) divergent.
Thus it is important to use chiral valence fermions in defining the currents on the lattice.
In our numerical calculation we use a discretized version of \Eq{eq:CLR4dintegral}.
We impose an upper limit $M$ in the summation over momenta, along the lines of \Eq{Ccontcut}, with $M< \pi/a$.
For non-zero valence mass $m_v$, one expects the dependence of $\CLR(m_v;M)$ to contain a quadratically divergent
term $\sim m_v^2/a^2$; this dependence should vanish when $m_v\to0$.

\subsection{Lattice \label{sec:ch_latt}}

We have seen that we require an axial current that is exactly conserved in the chiral limit.
In our previous work \cite{DeGrand:2016htl}, we chose to use overlap fermions for this purpose.
Here, for reasons of economy, we work with staggered fermions.
The conserved U(1) vector current is (see for example Ref.~\cite{DeGrand:2006zz})
\be
  V_\mu(x) = \frac{\eta_\mu(x)}{2}
  \left( \bar\chi(x) U_\mu(x) \chi(x+\hat\mu)
         + \bar\chi(x+\hat\mu) U^\dagger_\mu(x) \chi(x) \right) ,
\label{stagV}
\ee
while the partially conserved $\textrm{U}(1)_\epsilon$ axial current is
\be
  A_\mu(x) = \frac{\eta_\mu(x)\epsilon(x)}{2}
  \left( \bar\chi(x) U_\mu(x) \chi(x+\hat\mu)
         - \bar\chi(x+\hat\mu) U^\dagger_\mu(x) \chi(x) \right) \ .
\label{stagA}
\ee
The sign factors are, as usual,
\be
  \eta_1(x) = 1 \ , \quad
  \eta_2(x) = (-1)^{x_1} \ ,\quad
  \eta_3(x) = (-1)^{x_1+x_2} \ ,\quad
  \eta_4(x) = (-1)^{x_1+x_2+x_3} \ ,
\label{etamu}
\ee
and $\epsilon(x)=(-1)^{x_1+x_2+x_3+x_4}$.
These currents correspond to the nearest-neighbor staggered action.
The vector current satisfies the continuity equation
\be
  \sum_\mu \partial^-_\mu V_\mu(x)
  = \sum_\mu\Big(V_\mu(x) - V_\mu(x-\hat\mu)\Big) = 0 \ ,
\label{conservation}
\ee
and the axial current satisfies a similar continuity equation
in the massless limit.

We calculate the current--current correlation function $\PiLR(q)$ with the staggered currents (\ref{stagV}), (\ref{stagA}) exactly as was done with the overlap currents in Ref.~\cite{DeGrand:2016htl}.
Writing the chiral currents,
\bea
J^L_{\mu a}&=& V_{\mu a}-A_{\mu a}\,,\nonumber\\
J^R_{\mu a}&=& V_{\mu a}+A_{\mu a}\,,
\eea
we define the lattice correlator,
\be
\frac12\delta_{ab}\Pi_{\mu\nu}^{\textrm{lat}}(q) = -\frac14 a^4\sum_x \,e^{iqx}\, \left\langle J_{\mu a}^L(x) J_{\nu b}^R(0)\right\rangle\ .
\label{eq:latPiLR}
\ee
The factor of $\frac14$ corrects for the summation over the four tastes inherent in the staggered field.
The $\frac12$ is the normalization of isospin generators, $\tr T_aT_b=\frac12 \delta_{ab}$.
As usual,%
\footnote{Our lattice has $N_s^3\times N_t$ sites and we define $L_\mu=N_\mu a$.}
$q_\mu = (2\pi/L_\mu)n_\mu$ for periodic boundary conditions with period $L_\mu$, where $-N_\mu/2<n_\mu\le N_\mu/2$.
The desired $\PiLR(q)$ is the transverse piece of the correlation function.  We use a lattice definition,
\be
\PiLR(q) = \frac{\sum_{\mu\nu}P^T_{\mu\nu}(q)\Pi^{\textrm{lat}}_{\mu\nu}(q)}{3(\qhat^2)^2},
\label{eq:transverse}
\ee
that uses the lattice projector,
\be
P^T_{\mu\nu}(q) = \qhat^2 \delta_{\mu\nu} - \qhat_\mu \qhat_\nu\ ,
\ee
where $\qhat_\mu=(2/a)\sin (q_\mu a/2)$
and $\qhat^2\equiv\sum_\mu\qhat_\mu^2$.
Any lattice projector of course introduces lattice artifacts.
Their effects will vanish when we take the continuum limit.

\section{Lattice calculations \label{sec:numerical}}

Our results are based on the measurement of $\PiLR(q)$ on ten ensembles, to be described below.
In this section we give the methods applied to each ensemble separately, using one ensemble for illustration.
First we present our smearing procedure and describe its effect on the taste spectrum, as a function of the valence mass $m_v$.
Then we present $\PiLR(q)$ and compare it to a physical model, the minimal hadron approximation.
Finally, using values of $f_\pi$ obtained from valence spectroscopy, we calculate $\CLR(m_v)$ from $\PiLR$ and take the chiral limit, $m_v\to0$, for the staggered valence fermions on each ensemble.
We combine the ensemble results in Sec.~\ref{sec:limits}, giving a fit for $\CLR$ where the independent variables are the lattice spacing and the masses of the two sea fermions, measured in each ensemble; this fit produces continuum and chiral limits for $\CLR$.

\subsection{Overview of the ensembles \label{sec:ensembles}}

We base this study on 10 ensembles created for previous work.
They were generated with a lattice action containing an SU(4) gauge field coupled to dynamical fermions in both the fundamental \4~and the two-index antisymmetric \6~representation of SU(4), with two Dirac flavors of each.
For the fermions we use a Wilson-clover action, with normalized hypercubic (nHYP) smeared gauge links~\cite{Hasenfratz:2001hp,Hasenfratz:2007rf}.
The clover coefficient is set equal to unity for both fermion species~\cite{Bernard:1999kc,Shamir:2010cq}.
The pure gauge part of the action is the usual plaquette term plus an nHYP dislocation-suppressing (NDS) term, a smeared action designed to reduce gauge-field roughness that would create large fermion forces in the molecular dynamics evolution~\cite{DeGrand:2014rwa}.

The ensembles have lattice volumes of $N_s^3\times N_t=16^3\times32$ and $24^3\times48$.
They were generated and first used to calculate the meson spectrum of this model \cite{Ayyar:2017qdf}.
Twelve ensembles with volume $16^3\times32$ were then used to calculate the baryon spectrum, including mixed-representation ``chimera" baryons \cite{Ayyar:2018zuk}; a subset of these was used in calculating matrix elements of baryonic currents, in connection with partial compositeness  \cite{Ayyar:2018glg}.
In the present calculation of $\CLR$ we use nine of the $16^3\times32$ ensembles and one ensemble of volume $24^3\times48$.

Details relating to the ensembles appear in Appendix~\ref{app:ensembles}\@.
Table~\ref{table:ensembles} gives the values of the bare parameters used to generate the ensembles: the gauge coupling $\beta$ and the two hopping parameters, $\kappa_4$ and~$\kappa_6$, for the two fermion types.
We connect them to physics by the calculated values of the flow parameter $t_0/a^2$ and the fermion masses $m_4a$ and $m_6a$, obtained from the axial Ward identities (AWI)\@.
We use $t_0$ for setting the scale \cite{Ayyar:2017qdf} (for example, $t_0\simeq (1.4\ \textrm{GeV})^{-2}$ in QCD), whence we define the dimensionless lattice spacing
$\hat a=a/\sqrt{t_0}$ for each ensemble and the dimensionless AWI masses $\hat m_4=m_4\sqrt{t_0}$ and $\hat m_6=m_6\sqrt{t_0}$, as well as $\hCLR = \CLR\,t_0^2$\@.
We will use these quantities in taking continuum and chiral limits.
The measured values for these quantities appear in Table~\ref{table:fermion_masses}\@.
We refer the reader to Refs.~\cite{Ayyar:2017qdf,Ayyar:2018zuk} for more information about the ensembles.

\subsection{Staggered valence fermions: smearing and the taste spectrum \label{sec:taste}}

As discussed above, in order to maintain
an exact chiral symmetry,
we calculate valence propagators with a staggered fermion action.
We use a nearest-neighbor action, improved by smearing the SU(4) gauge field before promoting it to the sextet representation.
As in the Wilson action of the sea fermions, the smearing is an nHYP scheme;  the  smearing parameters are the same for the valence and the sea fermions.
In an effort to reduce lattice artifacts, we smear more than once to derive the valence action.
We calculate the valence propagators
for seven values of the valence mass for all the ensembles, with $0.01 \le m_va\le 0.05$.

\begin{figure}[t]
\begin{center}
\includegraphics[width=0.4\columnwidth,clip]{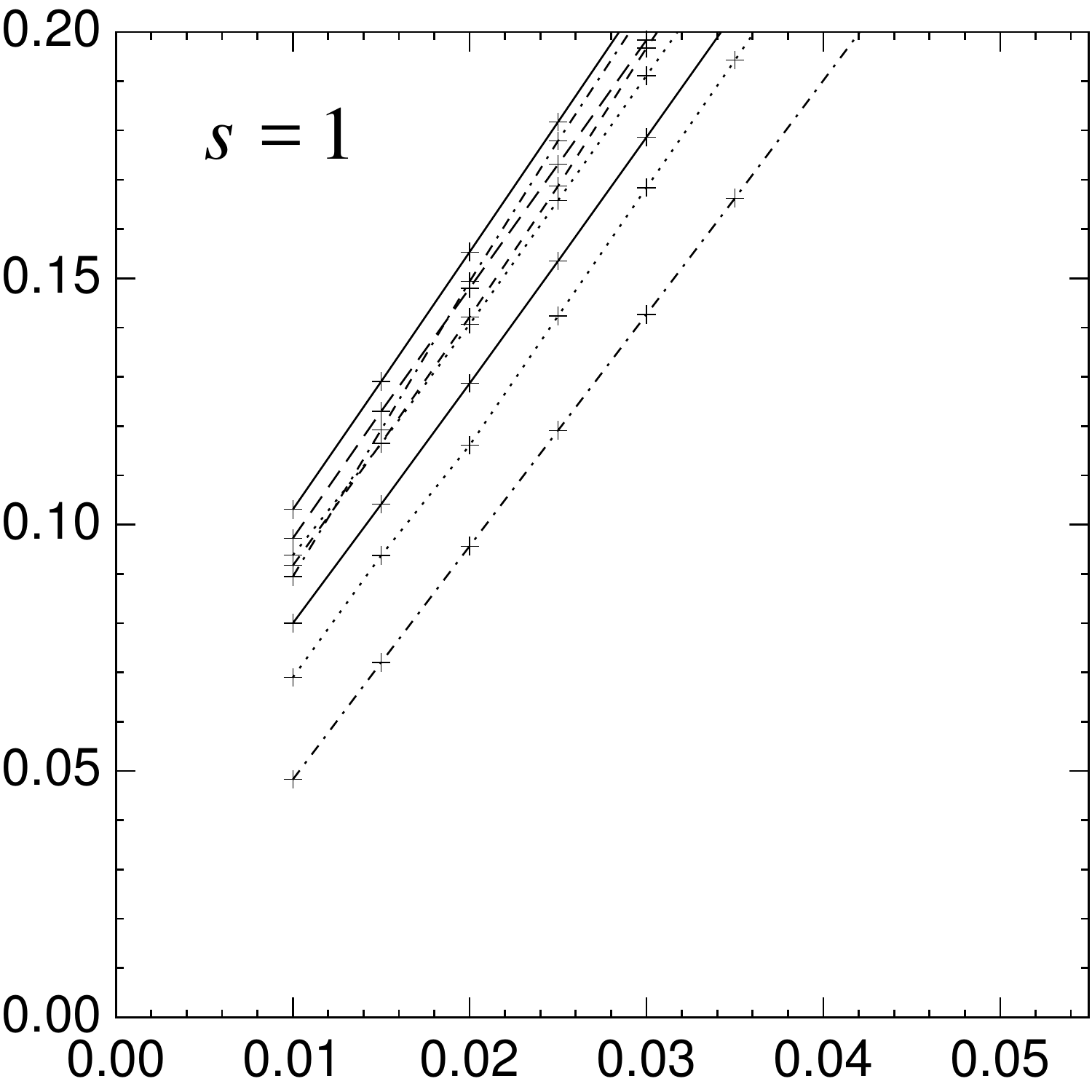}
\includegraphics[width=0.4\columnwidth,clip]{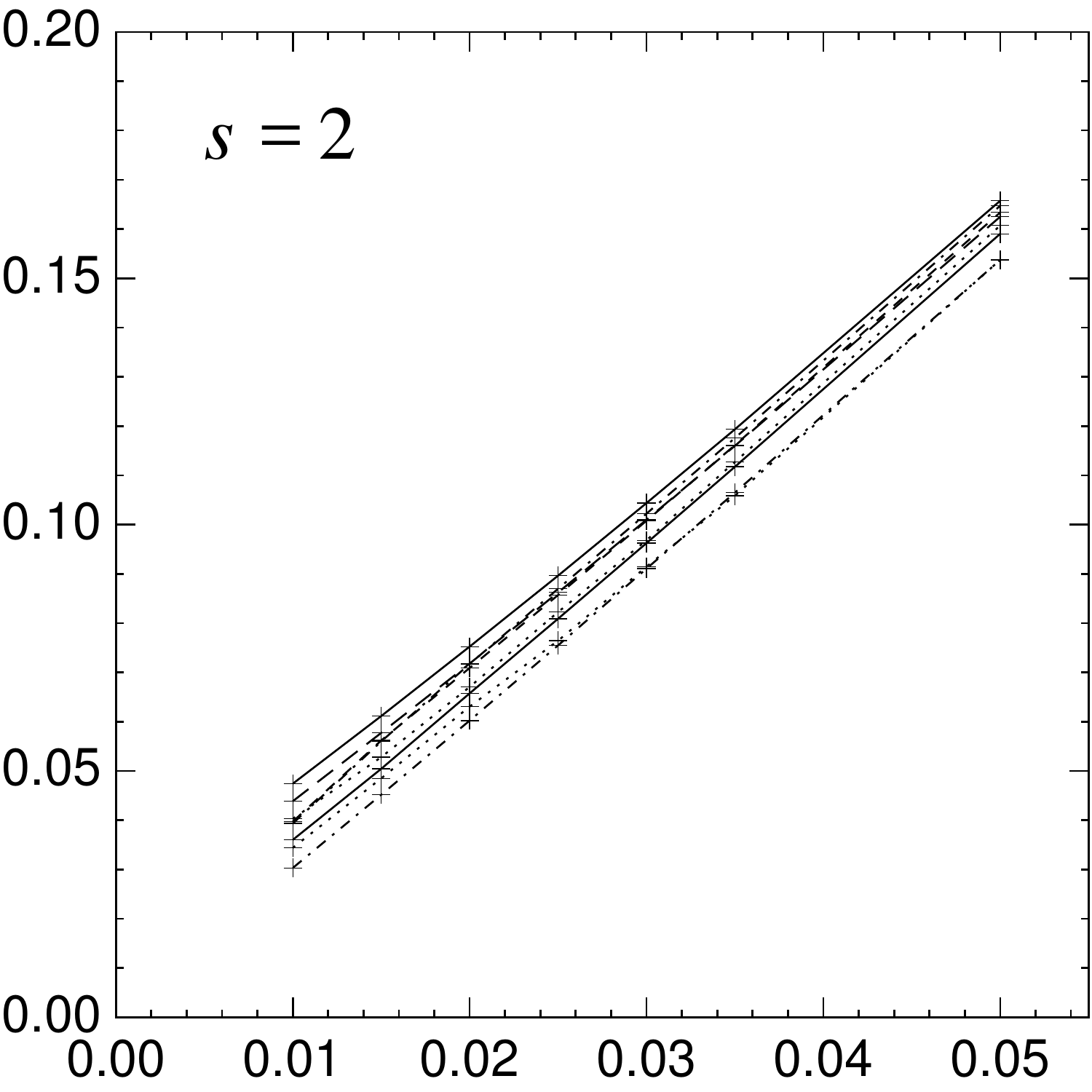}
\includegraphics[width=0.4\columnwidth,clip]{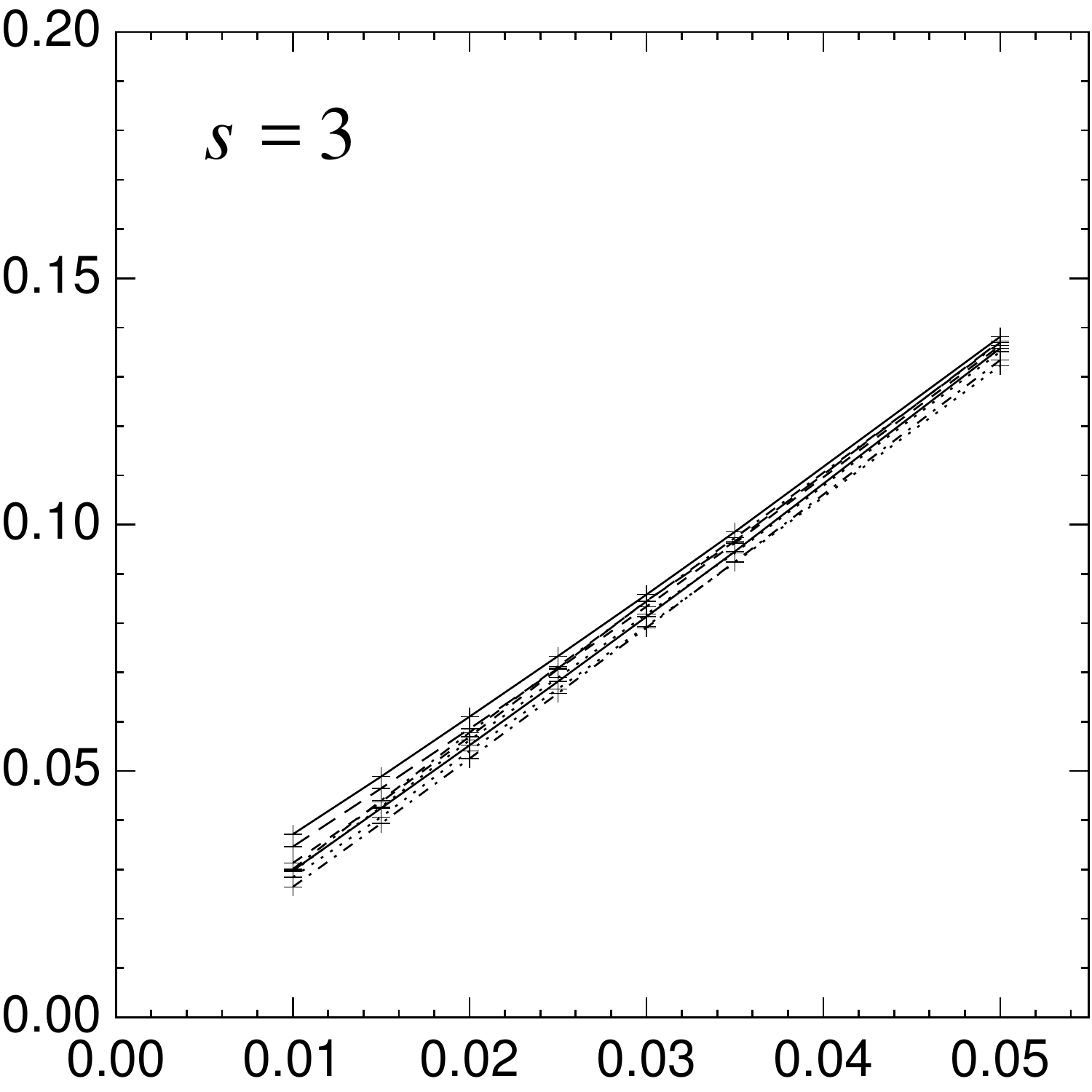}
\includegraphics[width=0.4\columnwidth,clip]{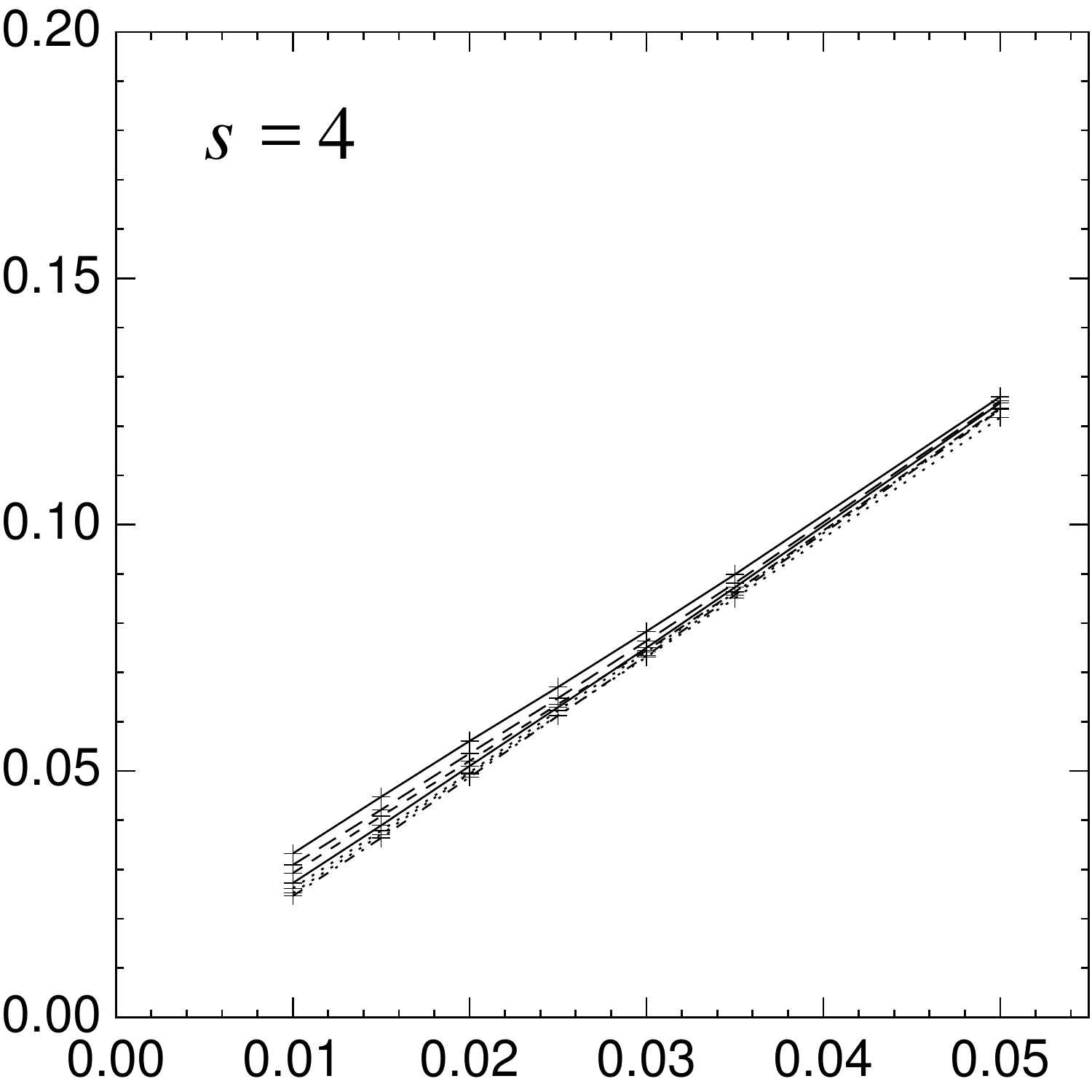}
\end{center}
\caption{Squared mass $(m_\pi a)^2$ of the valence Goldstone pion and
the seven additional pion multiplets,
constructed from the staggered valence fermions, in ensemble 2.
The abscissa is $m_va$, the valence fermion mass.
The four plots correspond to $s=1$, 2, 3, and 4 levels of smearing, carried out on the gauge field configurations before calculating the fermion propagators.
The lines connect points of each multiplet.
The Goldstone pion is always the bottom curve.
\label{fig:taste}}
\end{figure}
Figure~\ref{fig:taste} shows the splitting of the eight pion multiplets (that differ in their internal taste structure \cite{Golterman:1985dz,Lee:1999zxa}) in ensemble~2, calculated with one to four levels of smearing.
Going from one to two levels of smearing dramatically reduces the taste splittings, while additional smearing produces only marginal improvement.
The slope of the Goldstone pion's mass squared, which is a low energy constant, also changes only slightly after two smearings.
For production, we stop at two smearing levels, for two reasons.
\begin{enumerate}
\item We eventually combine the ensembles to arrive at a continuum extrapolation.
There is no reason to eliminate lattice artifacts before then; we only require them to be small enough that they do not hamper the extrapolation.
\item Repeated smearing increases the range of the action.
This may introduce a sensitivity to the finite volume.
\end{enumerate}

\subsection{Minimal hadron approximation and $f_\pi$ \label{sec:fpi}}

\begin{figure}
\begin{center}
\includegraphics[width=0.6\columnwidth,clip]{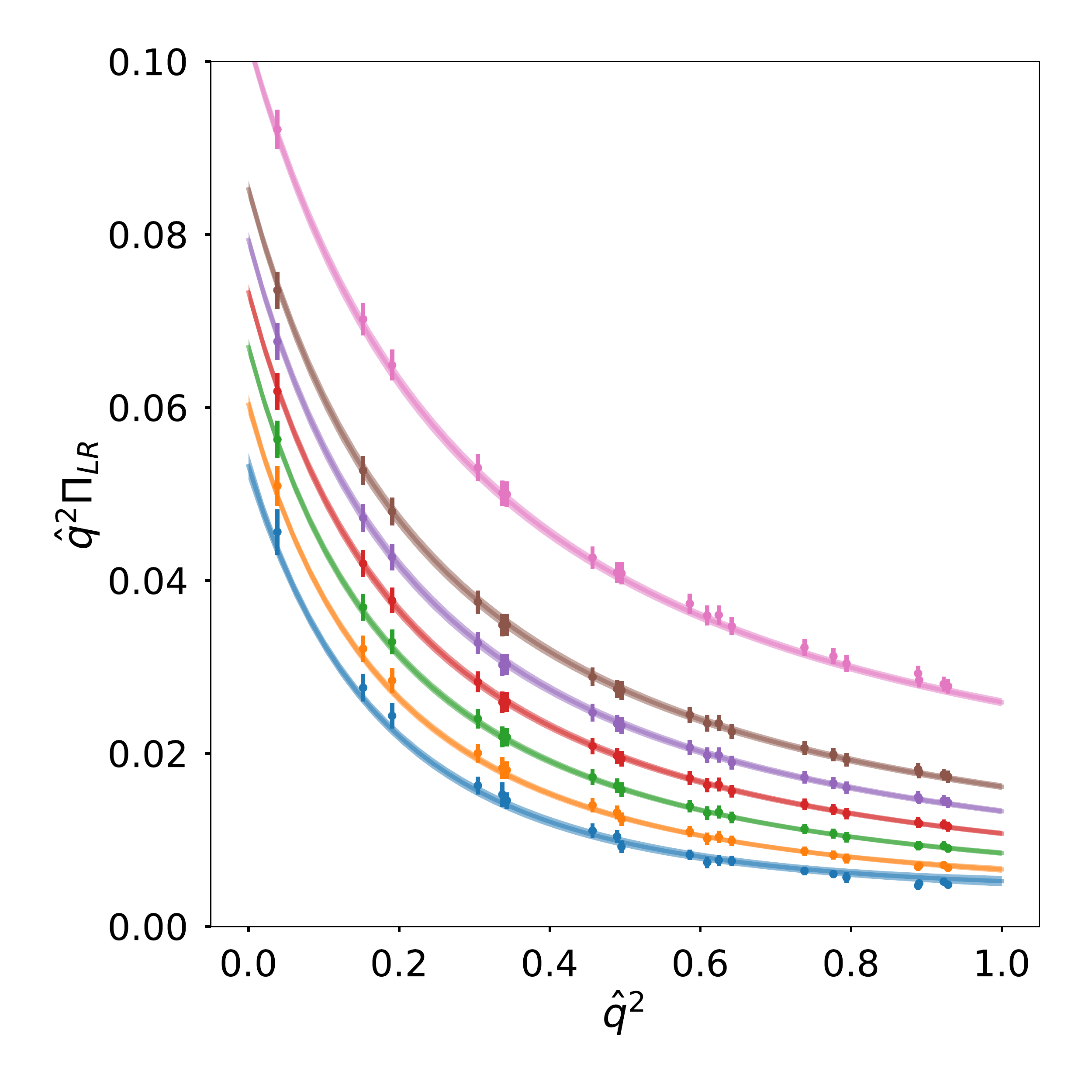}
\end{center}
\caption{
The transverse part $\PiLR$ of the current--current correlation function, plotted against
$\qhat^2=\sum_\mu \qhat_\mu^2$, in ensemble 2.
We plot the data and the fit to the MHA for the seven values of the valence mass $m_v=0.05$, 0.035, 0.03, 0.025, 0.02, 0.015, 0.01, top to bottom.
\label{fig:MHA}}
\end{figure}

The minimal hadron approximation (MHA) provides a physically appealing interpretation of $\PiLR(q)$ at low momentum.
The MHA supposes that $\PiLR(q)$ may be modeled by retaining poles from the pion and the lightest states in the axial and vector channels (the analogues of the $a_1$ and $\rho$ mesons in QCD, respectively):
\be
  \PiLR(q) \approx
  \frac{f_\pi^2}{q^2} + \frac{f_{a_1}^2}{q^2+m_{a_1}^2}
  - \frac{f_\rho^2}{q^2+m_\rho^2} \,.
\label{MHA}
\ee
The first term  comes from applying the transverse projection to the $f_\pi$-dependent term in the current correlator and does not depend the pion mass
(see Ref.~\cite{DeGrand:2016htl}).
We conducted fits to this function on each ensemble.
For illustration, Fig.~\ref{fig:MHA} shows $\qhat^2 \PiLR(\hat{q})$ calculated and fit for each valence mass $m_v$ for ensemble 2.
We fit to data in the range $\qhat^2<0.5$, since the MHA is only expected to hold at low momentum.
The figure shows that the fit works well at least up to $\qhat^2=1$.

We can compare the value of $f_\pi$ emerging from the fit to \Eq{MHA} to the spectroscopic value of $f_\pi$, which we calculate from the correlation function of the pseudoscalar density,
\be
  -a^3\sum_{\bf x}\svev{P^a({\bf x},t) P^b({\bf0},0)} 
  = 4\,\frac{f_\pi^2 m_\pi^3}{8 m_q^2}\,\delta^{ab} e^{-m_\pi t} ,
\label{fpiPP}
\ee
where $a,b$ are isospin indices and (again) the factor of 4 is for the four staggered tastes.
In our previous work~\cite{DeGrand:2016htl}, we found that these two values for $f_\pi$ match closely, for both ensembles studied there.
We compare the two determinations of $f_\pi$ in Fig. 3, which is typical
of all our ensembles.  The discrepancies are a combination of the
limitations of the three-pole ansatz and scaling violations.
We also show $f_{P6}$, the decay constant of the dynamical sextet Wilson pions, calculated in Ref.~\cite{Ayyar:2017qdf}.
Its rough agreement with the valence values suggests that
the added discretization error due to the use of a mixed action,
coming from the difference in the finite parts of renormalization constants,
is not large.

In Ref.~\cite{DeGrand:2016htl} we calculated $\CLR$ by integrating the MHA fit as well as by direct summation of $\PiLR$ in momentum space.
The MHA fits worked surprisingly well even in the UV regime, allowing us to use them for the complete momentum integral.
In the current work, we find that the fits do not work well at momenta outside the plot in Fig.~\ref{fig:MHA}, perhaps because of large discretization effects in the staggered formalism.
Hence, in Sec.~\ref{sec:CLR} below, we discard the MHA fits in favor of direct summation.
Moreover, we use the spectroscopic value of $f_\pi$ rather than that which emerges from the MHA fit.

\begin{figure}
\begin{center}
\includegraphics[width=0.6\columnwidth,clip]{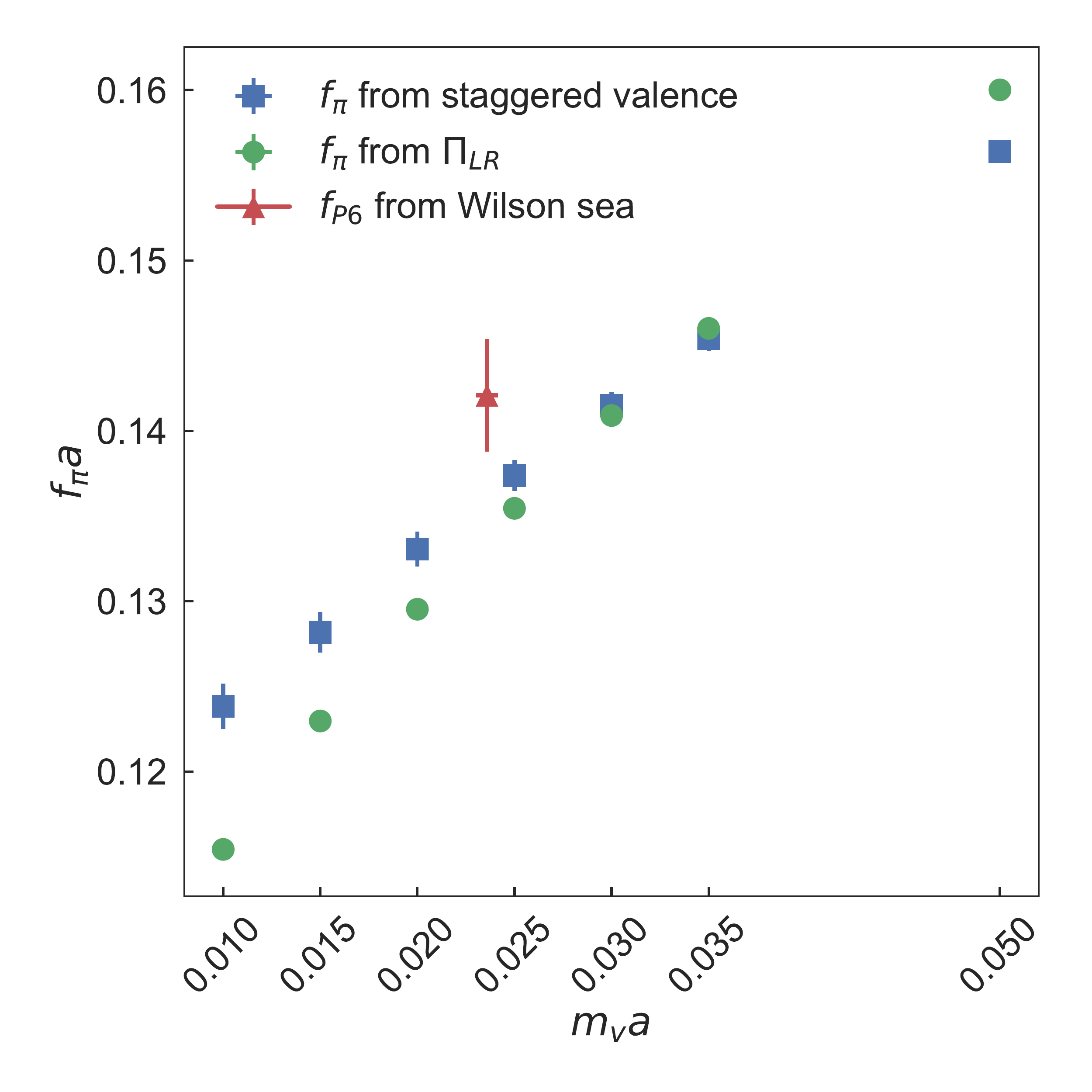}
\end{center}
\caption{
Comparison of $f_\pi$, the decay constant of the sextet valence pions, calculated from the MHA fit to $\PiLR(q)$, to that calculated directly from \Eq{fpiPP} in ensemble 2.
The abscissa is the mass of the valence fermions.
The error bars of the green circles are smaller than the symbols.
Also shown is $f_{P6}$, the decay constant of the sextet pions in the Wilson sea.
For this, the abscissa is the sextet AWI mass measured in the ensemble.
\label{fig:fpi}}
\end{figure}

\subsection{$\CLR(m_v)$ and the valence chiral limit \label{sec:CLR}}

To calculate $\CLR(m_v)$ on each ensemble, we follow the method of Ref.~\cite{DeGrand:2016htl}.
We compute the four-dimensional integral in \Eq{eq:CLR4dintegral} as direct summation of $\PiLR(q_\mu)$ in discrete momentum space.
The discrete version of \Eq{eq:CLR4dintegral} is
\be
\CLR=\frac{16\pi^2}V \sum_{q_\mu}\PiLR(q_\mu),
\label{summation}
\ee
where $V=L_s^3L_t$.
A sum over the entire Brillouin zone of the lattice would run over momenta $|q_\mu a|<\pi$, in steps of $\Delta q_\mu a=2\pi/N_\mu$.
In the present analysis with staggered fermions, such a sum would include  contributions from the Brillouin zone face that represent currents with various (adjoint) taste structures in the continuum limit.
We limit the summation in \Eq{summation} to the ``reduced Brillouin zone,''
\begin{align} \label{cutoff}
|q_\mu a|\leq q_{\textrm{max}}a=\pi/2,
\end{align}
or to a subset thereof (see below).
This retains only the taste-singlet vector current and the partially conserved axial current, which contains $\gamma_5\xi_5$ in Dirac-taste space.
With any prescription for the momentum sum, the lattice defines the divergent part of $\CLR(m_v)$ when $m_v$ is nonzero, proportional to $m_v^2/a^2$\@.

Equation~(\ref{summation}) requires special care at the origin of momentum space.
In the first place, the transverse projection (\ref{eq:transverse}) is undefined at $q=0$;
in the second place, $\PiLR$ in the continuum contains a kinematical pole at $q=0$ (as is seen, for instance, in the MHA).
Near zero, then, we approximate the continuum function as
\be
\PiLR(q_\mu)\simeq \frac{f_\pi^2}{q^2}+p,
\label{pole}
\ee
where we take the value of $f_\pi$ from the spectroscopic data and estimate the pedestal $p$ from $m$ neighboring momenta  (see Ref.~\cite{DeGrand:2016htl}),
\begin{equation}
p=\frac1m\sum_{a=1}^m\left(\PiLR(q^a)-\frac{f_\pi^2}{(q^a)^2}\right).
\label{neighbors}
\end{equation}
That is, $p$ is the average of the discrete $\PiLR$ in the neighboring cells, minus the pole term on those cells.
The contribution of $q=0$ to the discrete summation~(\ref{summation}) is then
\be
A\frac{L_s^2}V  \,f_\pi^2+\frac{16\pi^2}V p,
\ee
where $A\circeq22.5095963$ is a geometric factor calculated for the aspect ratio of our lattices, $L_t/L_s=2$.

\begin{figure}
\begin{center}
\includegraphics[width=.6\columnwidth,clip]{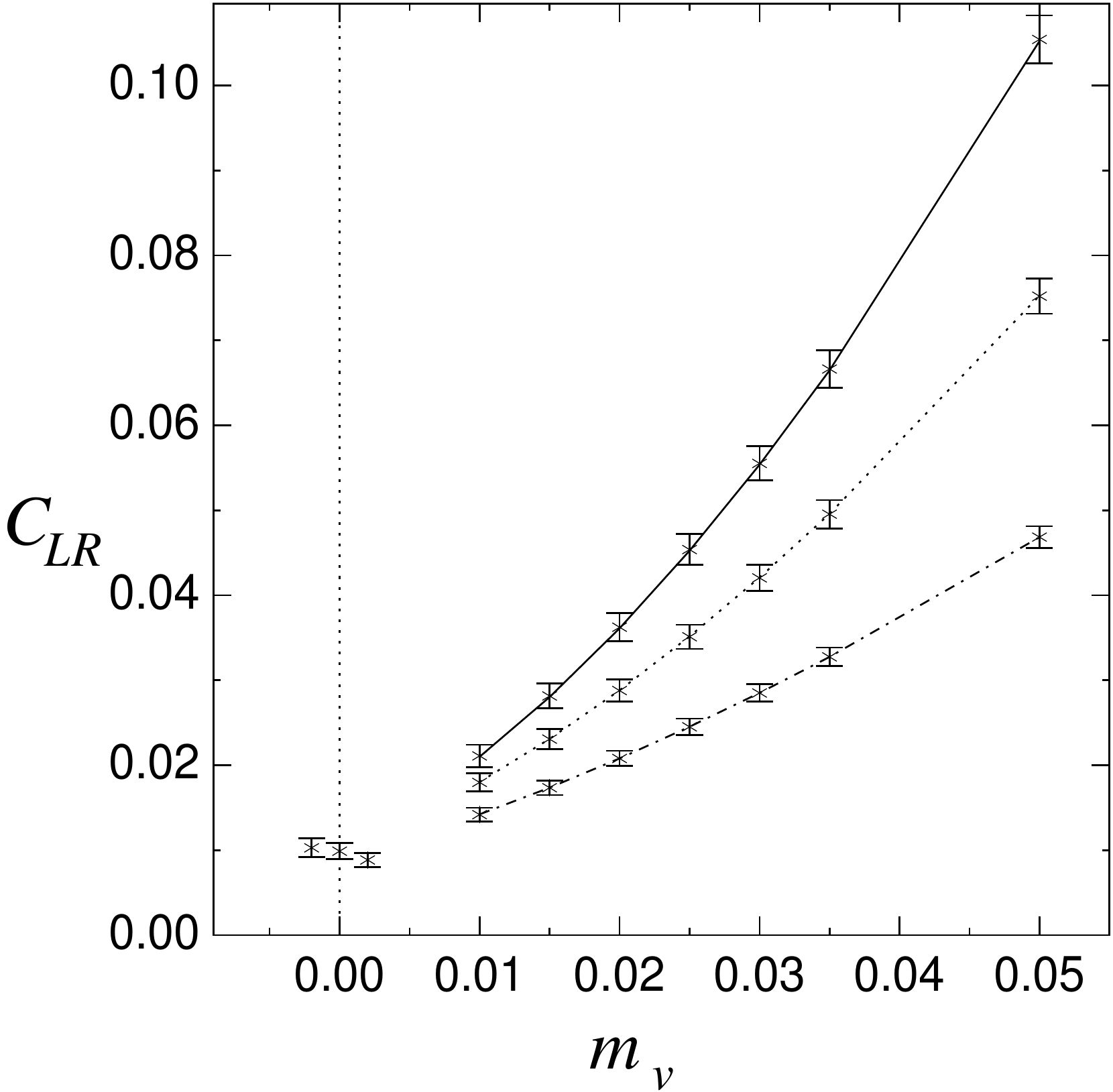}
\end{center}
\caption{
The low-energy constant $\CLR$ plotted against the valence mass $m_v$ (both in lattice units) in ensemble 2.
The three sets of data and the corresponding curves come from three different high-momentum cutoffs in the summation~(\ref{summation}).
The points plotted at $m_v=0$ (slightly shifted) are the cubic extrapolations of the data to zero.
The correlated errors are calculated from single-elimination jackknife.
\label{fig:CLRmv}}
\end{figure}

Figure~\ref{fig:CLRmv} shows the results for $\CLR(m_v)$ for ensemble 2, together with a cubic fit extrapolating to $m_v=0$.
The top set of points corresponds to the summation (\ref{summation}) over the entire reduced Brillouin zone, \Eq{cutoff}.
The extrapolation reaches $m_v=0$ at the point plotted slightly to the left of zero.
To examine the effect of the UV cutoff, we drop the highest momenta in the reduced Brillouin zone.
Because ensemble 2 has volume $16^3\times32$, this amounts to decreasing
\begin{align}
q_{\textrm{max}}a \rightarrow \frac{\pi}{2}-\frac{2\pi}{N_s} = \frac{3\pi}{8}.
\end{align}
The resulting values  for $\CLR(m_v)$ give the middle data set and curve in Fig.~\ref{fig:CLRmv}, extrapolated to the point plotted exactly at $m_v=0$.
As a third variation, we decrease the cutoff further,
$q_{\textrm{max}}a \rightarrow {\pi}/{4}$,
which gives the bottom data set and curve, extrapolated to the point plotted slightly to the right of zero.
The result of varying the UV cutoff is as expected: There is a marked change for nonzero valence mass, while $\CLR(0)$ is unchanged given the error bar.

The corresponding results for all the ensembles are quite similar.
The $\chi^2/\textrm{dof}$ of each fit is in the neighborhood of 0.3 for all ensembles.
We list the extrapolated $\CLR(0)$ in Table~\ref{table:CLR0}\@.
The use of a cubic for this fit, though arbitrary, gives an extrapolation that is stable under changes in the fitting polynomial.
Chiral perturbation theory, which gives formulas for many quantities, is not helpful here.
It can only treat $\PiLR(q)$ at low $q^2$, and hence it does not give a useful formula for $\CLR(m_v)$.

The extrapolation to $m_v=0$ also offers a means to test for sensitivity to finite-volume effects in $\CLR$.
Regarding mesonic quantities based on the sea fermions, the possibility of finite-volume contamination was addressed at length in Ref.~\cite{Ayyar:2017qdf} and dismissed.
The new ingredient in this work is the light valence fermions.
Finite-volume effects would of course be strongest for the smallest values of $m_v$.
A simple test is thus to drop the smallest values and examine anew the extrapolation of $\CLR$ to $m_v=0$.
We find that dropping the two smallest masses, $m_v=0.01$ and~0.015, shifts each extrapolation by less than its error bar.
Given the small $\chi^2$ of each fit, this is not surprising.
Moreover, the fits show no sign of requiring a term proportional to $1/(m_v\sqrt{V})$, an expected consequence of zero modes in a partially quenched theory.
See Ref.~\cite{DeGrand:2016htl} for more discussion.

While chiral perturbation theory is not directly applicable to $\CLR$ itself,
one could argue that it can be used to understand the finite-volume correction
to $\CLR$, which is a large-distance quantity \cite{Aubin:2015rzx}.
In our mixed-action setup, finite-volume corrections are controlled by the mass $M_{vs}$ of the mixed (valence-sea) pion.
The latter satisfies the bound\footnote{%
  While Ref.~\cite{Bar:2010ix} considered a different mixed-action setup,
  the proof extends to all mixed-action theories.
}
\cite{Bar:2010ix}
\be
\label{eq:bound}
M_{vs}^2 \ge (M_{ss}^2 + M_{vv}^2)/2 \ .
\ee
where $M_{ss}$ and $M_{vv}$ are respectively the masses of the pure sea and valence pions.
For ensemble 2 we can read $M_{vv}$ off Fig.~\ref{fig:taste} and $M_{ss}$ from Table~\ref{table:fermion_masses} (the sextet pion mass $M_{P6}$).
We find that  $M_{vs}L_s\ge 3.9$ when the smallest valence mass is 0.01, and $M_{vs}L_s\ge4.4$ when it is 0.02.
In all other ensembles $M_{vs}L_s$ satisfies more stringent bounds.
The increase of $M_{vs}L_s$ from (at least) 3.9 to (at least) 4.4
thus lends further support to our conclusion that finite-volume effects
are negligible, within the precision of our calculation.

\begin{table}[t]
\centering
\setlength{\tabcolsep}{10pt} 
\begin{tabular}{cl}
\toprule
Ensemble	&  $\CLR a^4$  \\
\hline
2	& 0.0103(11)\\
3	& 0.0081(10)\\
4	& 0.0120(8) \\
5	& 0.0057(6) \\
6	& 0.0217(16)\\
8	& 0.0037(6) \\
9	& 0.0096(7) \\
11	& 0.0028(6) \\
12	& 0.0027(6) \\[5pt]
40	& 0.0036(3) \\
\toprule
\end{tabular}
\caption{
The low-energy constant $\CLR$, calculated via summation over the entire reduced Brillouin zone, extrapolated to zero valence mass.}
\label{table:CLR0}
\end{table}

\section{$\CLR$ in the chiral and continuum limit \label{sec:limits}}

The ultimate goal of this study is the value of the low-energy constant $\CLR$ in the continuum limit.
As discussed above, $\CLR$ is finite and therefore physical only in the chiral limit of the sextet fermion.
In the present partially quenched lattice calculation, unitarity requires a simultaneous chiral limit for the sea and valence sextet fermions.
Therefore, the sole free parameter in the continuum limit is the mass of the fundamental fermions.
In Sec.~\ref{sec:CLR} we took the chiral limit for the valence sextet fermions.
We now turn our attention to the continuum limit and chiral limit for the sextet sea fermions.

To conduct this joint limit, we consider the dimensionless product $\hCLR = \CLR t_0^2$.
We model our data with a simple linear function,
\be \label{eq:model_function}
\hCLR = p_0 + p_a \ha + p_6 \hm_6,
\ee
which neglects dependence on the fundamental fermion mass $\hm_4$.
Later we test the stability of the fit parameters against alternative models, e.g., including dependence on the fundamental fermion mass.
First, we construct jackknife correlation matrices among the lattice quantities $\CLR$, $m_4$, and $m_6$ on each ensemble.
We do not include correlations with the flow scale $t_0/a^2$
(i.e., with $\ha\equiv a/\sqrt{t_0}$ \cite{Ayyar:2017qdf}),
which has negligible error compared to the other quantities we extract.
We then conduct a correlated fit to \Eq{eq:model_function}, obtaining $p_0=0.028(4)$, $p_a=-0.021(4)$, and $p_6=0.16(3)$, for $\chi^2 = 9.2/7$~dof.

\begin{figure}
\begin{center}
\includegraphics[width=0.8\textwidth]{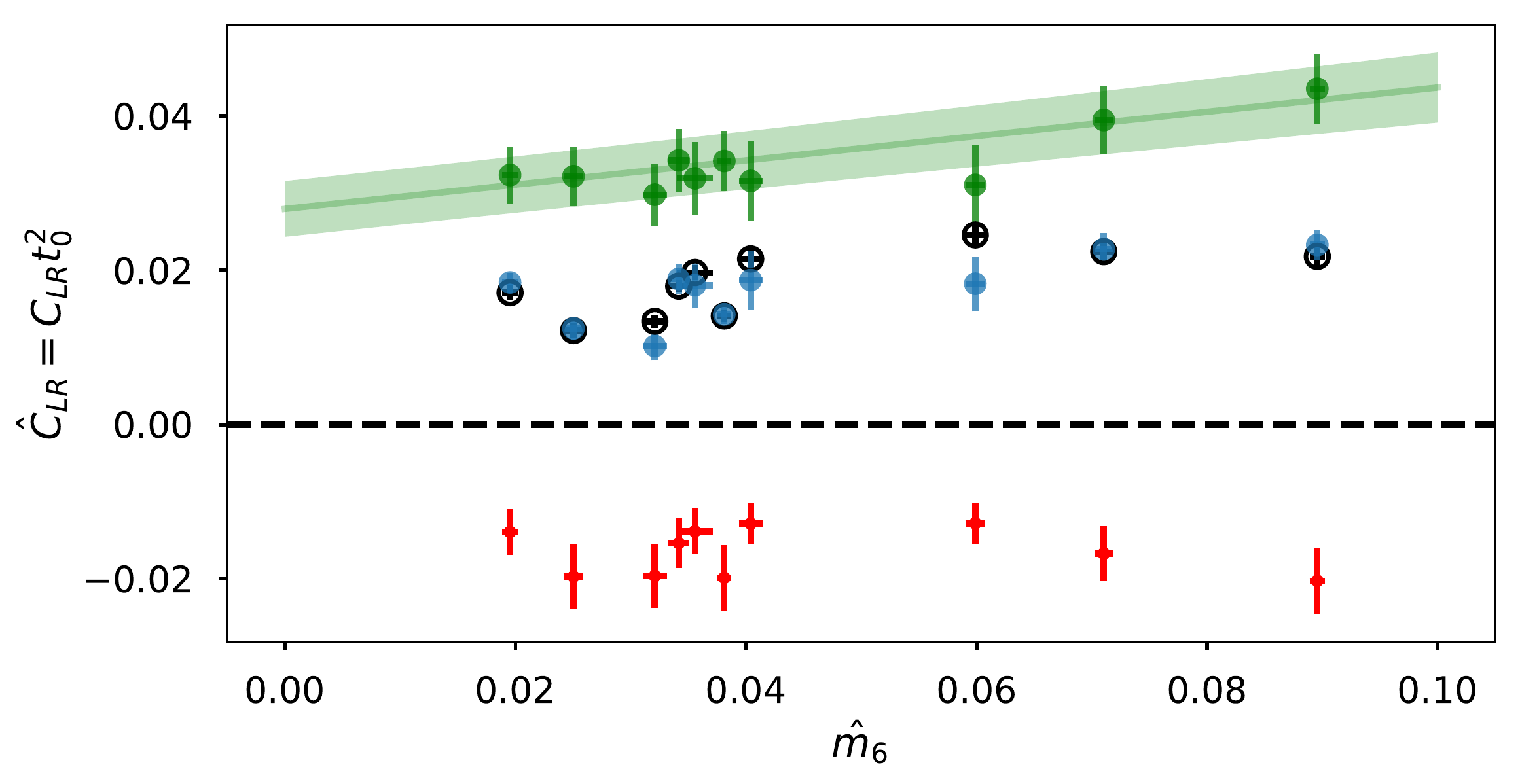}
\end{center}
\caption{
Fit of $\hCLR$ to \Eq{eq:model_function}.
The data appear in blue.
The best fit is in hollow black points.
The lattice artifact term $p_a \ha$ from \Eq{eq:model_function} identified by the fit appears in red.
The green points show the data minus the lattice artifact.
The smooth green band shows the continuum prediction, i.e., \Eq{eq:model_function} minus the $p_a \ha$ term.
\label{fig:minimal_fit}
}
\end{figure}
Figure~\ref{fig:minimal_fit} displays the result of the fit.
The hollow black points show the fit at the values of $\ha$ and $\hm_6$ of the individual ensembles; they follow the solid blue data points closely.
The lattice artifacts [$(p_a \ha)$ from \Eq{eq:model_function}] identified by the fit appear in red.
The green points show the data minus the lattice artifact.
According to the model, subtracting the artifacts from the full fit yields a linear function of $\hm_6$, which is displayed as a green band.
This band represents the continuum limit.

\begin{figure}
\begin{center}
\includegraphics[width=.3\textwidth]{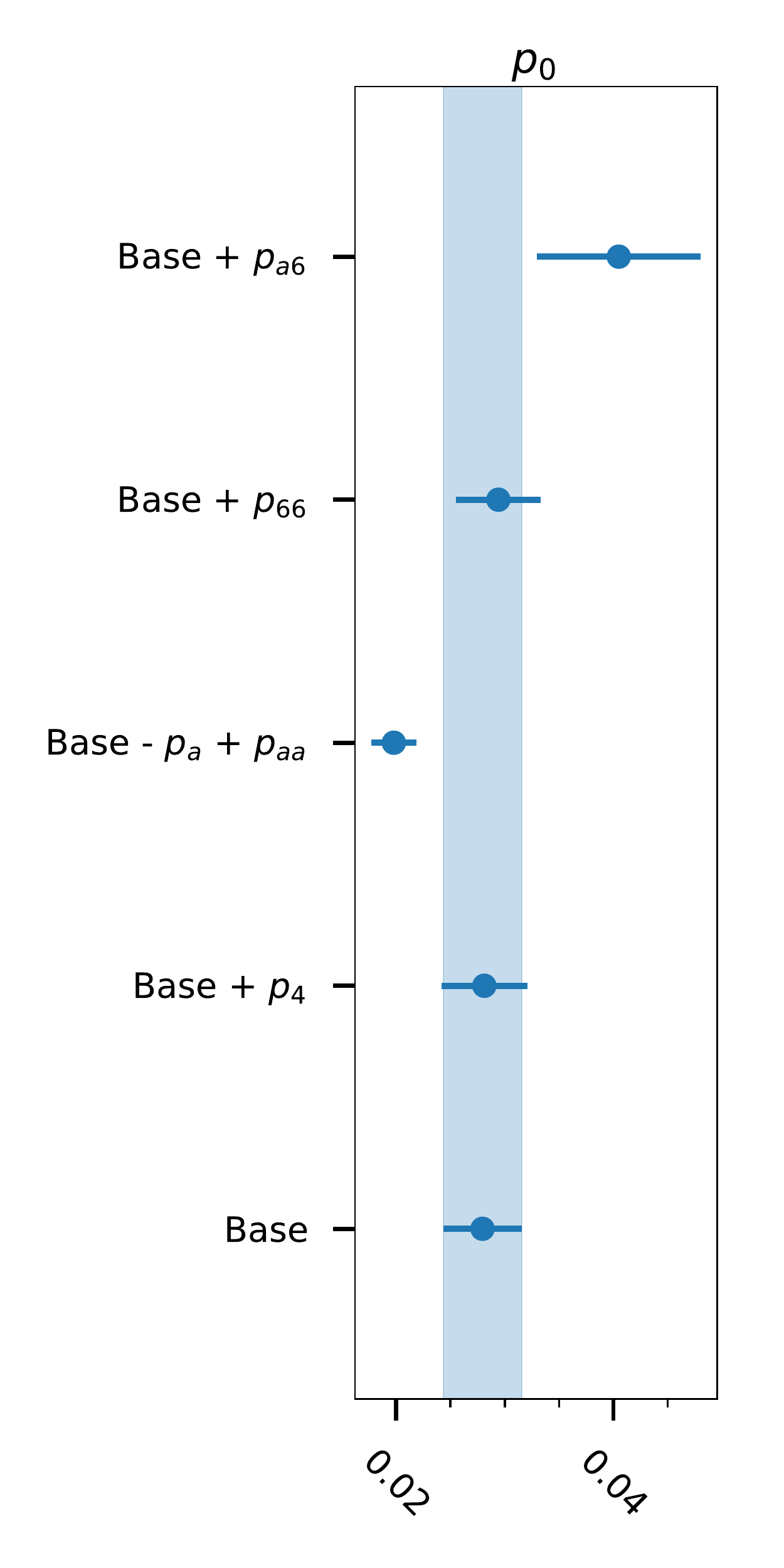}
\end{center}
\caption{
Stability of the fit parameter $p_0$ under model variations.
The shaded band corresponds to the fit to the base model of \Eq{eq:model_function}, shown in the bottom row.
The alternative models include additional terms shown in \Eq{eq:full_model}.
\label{fig:parameter_stability}
}
\end{figure}

It is significant that the fit, \Eq{eq:model_function}, works so well without including any dependence on $\hat m_4$, the mass of the fundamental fermions.
We have found before that the fundamental fermions have only a weak influence on quantities constructed from the sextet fermions~\cite{Ayyar:2017qdf,Ayyar:2018zuk}.
We can test the stability of our fit against the inclusion of an $\hm_4$ term, as well as higher-order terms in $\hm_6$ and $\ha$:
\be \label{eq:full_model}
\hCLR = p_0 + p_a \ha + p_6 \hm_6 +p_4 \hm_4 + p_{66}\hm_6^2 + p_{a6} \ha\hm_6 + p_{aa} \ha^2\ .
\ee
Figure~\ref{fig:parameter_stability} shows the stability of the best-fit result $p_0$ under the inclusion of these additional terms.  No significant discrepancy is seen; the largest deviation comes from the fit ``\mbox{$\textrm{Base}-p_a+p_{aa}$},'' which models the lattice-spacing dependence as quadratic instead of linear.  As can be seen from Fig.~\ref{fig:minimal_fit}, although the lattice-artifact contributions are significant, they vary little over the set of ensembles considered, which makes it difficult to determine whether linear or quadratic dependence on $\ha$ is a more appropriate description of our results.

The fit including the quadratic $p_{aa}$ yields a best-fit value of $p_0 = 0.020(2)$, as compared to the result $p_0 = 0.028(4)$ from the base fit.  Since we cannot reject either hypothesis, we conservatively adopt half of the difference in central values between these two fits as a systematic error, giving our final result
\be
\label{eq:final}
\hCLR = 0.024(4)_{\rm stat}(4)_{\rm sys}\ .
\ee
This is the chiral limit $m_v=m_6=0$ in both valence and sea-sextet masses, as well as the continuum limit, of $\hCLR$.

\section{Discussion \label{sec:discussion}}

In this paper we computed the low-energy constant $\CLR$ for the sextet representation.
We used staggered valence fermions to define the currents, instead of overlap fermions as in our previous work.
We indeed found this to be an economical way to impose exact chiral symmetry.
On the other hand, we find a strong dependence of $\CLR(m_v,M)$ on the cutoff momentum $M$, when the valence mass $m_v$ is nonzero.
When using overlap fermions, we could not resolve any dependence at all on $M$.
In any case, the divergent cutoff dependence disappears in the valence chiral limit, as expected for the quadratic divergence $m_v^2M^2\sim m_v^2/a^2$.

Taking the continuum limit as well as the chiral limit for both sea and valence sextet fermions we find $\hCLR =\CLR t_0^2 = 0.024(4)(4)$.
Within error, this result turns out to be independent of the mass of the fundamental fermions, for the mass range covered by our ensembles.  To simplify the lattice calculation we have considered a fermion content that is slightly different from Ferretti's model, so that our result is not directly applicable.
We expect, however, that our result for $\CLR$ will be numerically similar to that in Ferretti's actual model.

In Ferretti's model, the ratio $\CLR / f_{P6}^4$, where $f_{P6}$ is the sextet pseudoscalar decay constant, controls the contribution of the electroweak gauge bosons to the Higgs potential.  From \cite{Ayyar:2017qdf}, the value of $f_{P6}$ in the sextet chiral limit is $\sqrt{t_0} f_{P6} = 0.17(1)$.  Our determinations of $\CLR$ and $f_{P6}$ are found to be weakly correlated, so we can take the ratio and combine the errors in quadrature, finding the result
\be
\frac{\CLR}{f_{P6}^4} = 29(8)(5)\ .
\ee
In the notation of \cite{DelDebbio:2017ini}, this result translates to $\hat{c}_{LR} = \frac{1}{2} (3g^2 + g'{}^2) (\CLR / f_{P6}^4) = 19(5)(3)$.  Certain combinations of $\hat{c}_{LR}$ with other low-energy constants in the theory must be of order $10^{-2}$ to reproduce the observed properties of the Higgs boson, so this result implies some amount of fine-tuning in the theory.

It is interesting to compare our result with QCD.
Das et al.~\cite{Das:1967it} showed long ago that the electromagnetic mass splitting among pions is given by
\be
\label{eq:das}
m_{\pi^\pm}^2 - m_{\pi^0}^2
= \frac{3\alpha}{4\pi} \frac{\CLR}{f_\pi^2}\ .
\ee
Solving for $\CLR$ in terms of the experimental values gives, for QCD,
$\CLR \approx  0.012 \GeV^4$, and $\CLR / f_\pi^4 \approx 42$, where our
convention for the decay constant is $f_\pi \simeq 130 \MeV$.
Our result in the present model is thus comparable with the QCD value.  Expectations from large-$N_c$ \cite{Ayyar:2017qdf} are that $\CLR$ should scale as the dimension of the representation, {\em i.e.}, similar to $f_{P6}^2$.
The dimension of the antisymmetric representation is $N_c(N_c-1)/2$,
and so our smaller value for $\CLR / f_{P6}^4$ is in rough agreement
with such scaling, too.

We have recently computed the top-quark mass induced via partial compositeness in this model, finding it to be too small by several orders of magnitude.
This makes it extremely unlikely that Ferretti's model could generate a realistic top mass \cite{Ayyar:2018glg}.
However, the model might be rescued by introducing additional fermion species which are inert under all Standard Model interactions.
These will slow down the running, ultimately producing a nearly conformal but confining theory, which, in principle, might allow for large enhancement of the induced top mass.
Other models from the Ferretti-Karateev list \cite{Ferretti:2013kya} might give a more realistic top mass as well~\cite{Bennett:2017kga,Lee:2018ztv}.

\begin{acknowledgments}

Our calculations of staggered fermion propagators and currents were carried out with code derived from version 7.8 of the publicly available code of the MILC collaboration \cite{MILC}.
We thank Doug Toussaint for helpful correspondence and Thomas DeGrand for fruitful discussions.

Computations for this work were carried out with resources provided by the USQCD Collaboration,
which is funded
by the Office of Science of the U.S.\ Department of Energy.
This work was supported in part by the U.S.\ Department of Energy under grants DE-SC0010005 (Colorado) and DE-SC0013682 (SFSU) and by the Israel Science Foundation under grants no.~449/13 and no.~491/17 (Tel Aviv).
This manuscript has been authored by Fermi Research Alliance, LLC under Contract No. DE-AC02-07CH11359 with the U. S. Department of Energy, Office of Science, Office of High Energy Physics.

\end{acknowledgments}

\appendix

\section{The ensembles} \label{app:ensembles}
Table~\ref{table:ensembles} lists the ensembles used in this study.
The nine ensembles labelled 2--12 are lattices with volume $16^3\times32$, numbered as in Ref.~\cite{Ayyar:2018zuk} for easy reference; ensemble 40 has volume $24^3\times48$ (see Ref.~\cite{Ayyar:2017qdf}).
We have excluded ensembles (1, 7, 10) because large fluctuations in the observable $\CLR(m_v)$ indicate that the simulations are much too short.
Of our $24^3\times48$ ensembles, only ensemble 40 avoids this problem.

Table~\ref{table:fermion_masses} presents some physical properties of the ensembles.
These are the flow parameter $t_0$; the fermion masses from the axial Ward identities, rendered dimensionless as $\hat m_4=m_4\sqrt{t_0}$, and $\hat m_6=m_6\sqrt{t_0}$; and  the pseudoscalar meson masses, $\hat M_{P4}=M_{P4}\sqrt{t_0}$ and $\hat M_{P6}=M_{P6}\sqrt{t_0}$.

\begin{table}[t]
\centering
\setlength{\tabcolsep}{10pt} 
\begin{ruledtabular}
\begin{tabular}{clllc}
Ensemble	&  $\beta$ & $\kappa_4$ & $ \kappa_6$  & Configurations \\
\hline
2  &  7.25 &  0.13147 &  0.13395 &  71 \\
3  &  7.30 &  0.13117 &  0.13363 &  61 \\
4  &  7.30 &  0.13162 &  0.1334  &  71 \\
5  &  7.55 &  0.13    &  0.1325  &  85 \\
6  &  7.65 &  0.129   &  0.1308  &  50 \\
8  &  7.65 &  0.13    &  0.132   &  85 \\
9  &  7.75 &  0.128   &  0.131   &  85 \\
11 &  7.75 &  0.1295  &  0.1315  &  34 \\
12 &  7.85 &  0.129   &  0.1308  &  45 \\[5pt]
40 &  7.51 &  0.1307  &  0.1328  & 133 \\
\end{tabular}
\end{ruledtabular}
\caption{
	The ensembles used in this study.
	The lattice volumes are $16^3\times32$ (ensembles 2--12) and $24^3\times48$ (ensemble 40).
}
\label{table:ensembles}
\end{table}

\begin{table}[t]
\centering
\setlength{\tabcolsep}{10pt} 
\begin{ruledtabular}
\begin{tabular}{clllll}
Ensemble & $t_0/a^2$ & $\hat{m}_4$ & $\hat{m}_6$ & $\hat M_{P4}$ & $\hat M_{P6}$\\
\hline
2  &  1.135(9) &   0.028(1) &   0.025(1) &   0.305(9) &   0.303(8)\\
3  &   1.13(1) &  0.0345(8) &   0.032(1) &   0.340(5) &   0.340(9)\\
4  &  1.111(9) &  0.0228(6) &  0.0381(8) &   0.279(7) &  0.368(11)\\
5  &   1.85(2) &   0.050(1) &   0.034(1) &  0.429(10) &   0.375(9)\\
6  &  1.068(5) &   0.082(1) &  0.0896(8) &   0.514(8) &   0.576(9)\\
8  &   2.29(2) &   0.038(1) &   0.036(2) &   0.400(9) &  0.408(10)\\
9  &   1.56(1) &   0.108(1) &   0.071(1) &   0.636(7) &   0.538(8)\\
11 &   2.62(2) &   0.047(1) &   0.040(1) &  0.443(14) &  0.428(15)\\
12 &   2.67(2) &   0.060(1) &   0.060(1) &  0.505(13) &  0.529(17)\\[5pt]
40 &   2.26(2) &  0.0196(4) &   0.0194(9) &  0.278(4) &  0.291(10)\\
\end{tabular}
\end{ruledtabular}
\caption{
	Physical properties of the ensembles: Flow variable $t_0$, AWI masses $\hat m_r=m_r\sqrt{t_0}$, and pseudoscalar meson masses $\hat M_{Pr}=M_{Pr}\sqrt{t_0}$ for dynamical fermions in representations $r=4,6$.
}
\label{table:fermion_masses}
\end{table}

\bibliography{CLR2rep}

\begin{thebibliography}{43}%
\makeatletter
\providecommand \@ifxundefined [1]{%
 \@ifx{#1\undefined}
}%
\providecommand \@ifnum [1]{%
 \ifnum #1\expandafter \@firstoftwo
 \else \expandafter \@secondoftwo
 \fi
}%
\providecommand \@ifx [1]{%
 \ifx #1\expandafter \@firstoftwo
 \else \expandafter \@secondoftwo
 \fi
}%
\providecommand \natexlab [1]{#1}%
\providecommand \enquote  [1]{``#1''}%
\providecommand \bibnamefont  [1]{#1}%
\providecommand \bibfnamefont [1]{#1}%
\providecommand \citenamefont [1]{#1}%
\providecommand \href@noop [0]{\@secondoftwo}%
\providecommand \href [0]{\begingroup \@sanitize@url \@href}%
\providecommand \@href[1]{\@@startlink{#1}\@@href}%
\providecommand \@@href[1]{\endgroup#1\@@endlink}%
\providecommand \@sanitize@url [0]{\catcode `\\12\catcode `\$12\catcode
  `\&12\catcode `\#12\catcode `\^12\catcode `\_12\catcode `\%12\relax}%
\providecommand \@@startlink[1]{}%
\providecommand \@@endlink[0]{}%
\providecommand \url  [0]{\begingroup\@sanitize@url \@url }%
\providecommand \@url [1]{\endgroup\@href {#1}{\urlprefix }}%
\providecommand \urlprefix  [0]{URL }%
\providecommand \Eprint [0]{\href }%
\providecommand \doibase [0]{http://dx.doi.org/}%
\providecommand \selectlanguage [0]{\@gobble}%
\providecommand \bibinfo  [0]{\@secondoftwo}%
\providecommand \bibfield  [0]{\@secondoftwo}%
\providecommand \translation [1]{[#1]}%
\providecommand \BibitemOpen [0]{}%
\providecommand \bibitemStop [0]{}%
\providecommand \bibitemNoStop [0]{.\EOS\space}%
\providecommand \EOS [0]{\spacefactor3000\relax}%
\providecommand \BibitemShut  [1]{\csname bibitem#1\endcsname}%
\let\auto@bib@innerbib\@empty
\bibitem [{\citenamefont {Georgi}\ and\ \citenamefont
  {Kaplan}(1984)}]{Georgi:1984af}%
  \BibitemOpen
  \bibfield  {author} {\bibinfo {author} {\bibfnamefont {H.}~\bibnamefont
  {Georgi}}\ and\ \bibinfo {author} {\bibfnamefont {D.~B.}\ \bibnamefont
  {Kaplan}},\ }\href {\doibase 10.1016/0370-2693(84)90341-1} {\bibfield
  {journal} {\bibinfo  {journal} {Phys. Lett.}\ }\textbf {\bibinfo {volume}
  {145B}},\ \bibinfo {pages} {216} (\bibinfo {year} {1984})}\BibitemShut
  {NoStop}%
\bibitem [{\citenamefont {Dugan}\ \emph {et~al.}(1985)\citenamefont {Dugan},
  \citenamefont {Georgi},\ and\ \citenamefont {Kaplan}}]{Dugan:1984hq}%
  \BibitemOpen
  \bibfield  {author} {\bibinfo {author} {\bibfnamefont {M.~J.}\ \bibnamefont
  {Dugan}}, \bibinfo {author} {\bibfnamefont {H.}~\bibnamefont {Georgi}}, \
  and\ \bibinfo {author} {\bibfnamefont {D.~B.}\ \bibnamefont {Kaplan}},\
  }\href {\doibase 10.1016/0550-3213(85)90221-4} {\bibfield  {journal}
  {\bibinfo  {journal} {Nucl. Phys.}\ }\textbf {\bibinfo {volume} {B254}},\
  \bibinfo {pages} {299} (\bibinfo {year} {1985})}\BibitemShut {NoStop}%
\bibitem [{\citenamefont {Contino}(2011)}]{Contino:2010rs}%
  \BibitemOpen
  \bibfield  {author} {\bibinfo {author} {\bibfnamefont {R.}~\bibnamefont
  {Contino}},\ }in\ \href {\doibase 10.1142/9789814327183_0005} {\emph
  {\bibinfo {booktitle} {{Physics of the large and the small, TASI 09,
  proceedings of the Theoretical Advanced Study Institute in Elementary
  Particle Physics, Boulder, Colorado, USA, 1-26 June 2009}}}}\ (\bibinfo
  {year} {2011})\ pp.\ \bibinfo {pages} {235--306},\ \Eprint
  {http://arxiv.org/abs/1005.4269} {arXiv:1005.4269 [hep-ph]} \BibitemShut
  {NoStop}%
\bibitem [{\citenamefont {Bellazzini}\ \emph {et~al.}(2014)\citenamefont
  {Bellazzini}, \citenamefont {Cs{\'a}ki},\ and\ \citenamefont
  {Serra}}]{Bellazzini:2014yua}%
  \BibitemOpen
  \bibfield  {author} {\bibinfo {author} {\bibfnamefont {B.}~\bibnamefont
  {Bellazzini}}, \bibinfo {author} {\bibfnamefont {C.}~\bibnamefont
  {Cs{\'a}ki}}, \ and\ \bibinfo {author} {\bibfnamefont {J.}~\bibnamefont
  {Serra}},\ }\href {\doibase 10.1140/epjc/s10052-014-2766-x} {\bibfield
  {journal} {\bibinfo  {journal} {Eur. Phys. J.}\ }\textbf {\bibinfo {volume}
  {C74}},\ \bibinfo {pages} {2766} (\bibinfo {year} {2014})},\ \Eprint
  {http://arxiv.org/abs/1401.2457} {arXiv:1401.2457 [hep-ph]} \BibitemShut
  {NoStop}%
\bibitem [{\citenamefont {Panico}\ and\ \citenamefont
  {Wulzer}(2016)}]{Panico:2015jxa}%
  \BibitemOpen
  \bibfield  {author} {\bibinfo {author} {\bibfnamefont {G.}~\bibnamefont
  {Panico}}\ and\ \bibinfo {author} {\bibfnamefont {A.}~\bibnamefont
  {Wulzer}},\ }\href@noop {} {\bibfield  {journal} {\bibinfo  {journal} {Lect.
  Notes Phys.}\ }\textbf {\bibinfo {volume} {913}} (\bibinfo {year} {2016})},\
  \Eprint {http://arxiv.org/abs/1506.01961} {arXiv:1506.01961 [hep-ph]}
  \BibitemShut {NoStop}%
\bibitem [{\citenamefont {Ferretti}\ and\ \citenamefont
  {Karateev}(2014)}]{Ferretti:2013kya}%
  \BibitemOpen
  \bibfield  {author} {\bibinfo {author} {\bibfnamefont {G.}~\bibnamefont
  {Ferretti}}\ and\ \bibinfo {author} {\bibfnamefont {D.}~\bibnamefont
  {Karateev}},\ }\href {\doibase 10.1007/JHEP03(2014)077} {\bibfield  {journal}
  {\bibinfo  {journal} {JHEP}\ }\textbf {\bibinfo {volume} {03}},\ \bibinfo
  {pages} {077} (\bibinfo {year} {2014})},\ \Eprint
  {http://arxiv.org/abs/1312.5330} {arXiv:1312.5330 [hep-ph]} \BibitemShut
  {NoStop}%
\bibitem [{\citenamefont {Ferretti}(2014)}]{Ferretti:2014qta}%
  \BibitemOpen
  \bibfield  {author} {\bibinfo {author} {\bibfnamefont {G.}~\bibnamefont
  {Ferretti}},\ }\href {\doibase 10.1007/JHEP06(2014)142} {\bibfield  {journal}
  {\bibinfo  {journal} {JHEP}\ }\textbf {\bibinfo {volume} {06}},\ \bibinfo
  {pages} {142} (\bibinfo {year} {2014})},\ \Eprint
  {http://arxiv.org/abs/1404.7137} {arXiv:1404.7137 [hep-ph]} \BibitemShut
  {NoStop}%
\bibitem [{\citenamefont {Ferretti}(2016)}]{Ferretti:2016upr}%
  \BibitemOpen
  \bibfield  {author} {\bibinfo {author} {\bibfnamefont {G.}~\bibnamefont
  {Ferretti}},\ }\href {\doibase 10.1007/JHEP06(2016)107} {\bibfield  {journal}
  {\bibinfo  {journal} {JHEP}\ }\textbf {\bibinfo {volume} {06}},\ \bibinfo
  {pages} {107} (\bibinfo {year} {2016})},\ \Eprint
  {http://arxiv.org/abs/1604.06467} {arXiv:1604.06467 [hep-ph]} \BibitemShut
  {NoStop}%
\bibitem [{\citenamefont {Cacciapaglia}\ \emph {et~al.}(2019)\citenamefont
  {Cacciapaglia}, \citenamefont {Ferretti}, \citenamefont {Flacke},\ and\
  \citenamefont {Ser\^odio}}]{Cacciapaglia:2019bqz}%
  \BibitemOpen
  \bibfield  {author} {\bibinfo {author} {\bibfnamefont {G.}~\bibnamefont
  {Cacciapaglia}}, \bibinfo {author} {\bibfnamefont {G.}~\bibnamefont
  {Ferretti}}, \bibinfo {author} {\bibfnamefont {T.}~\bibnamefont {Flacke}}, \
  and\ \bibinfo {author} {\bibfnamefont {H.}~\bibnamefont {Ser\^odio}},\ }\href
  {\doibase 10.3389/fphy.2019.00022} {\bibfield  {journal} {\bibinfo  {journal}
  {Front. Phys.}\ }\textbf {\bibinfo {volume} {7}},\ \bibinfo {pages} {22}
  (\bibinfo {year} {2019})},\ \Eprint {http://arxiv.org/abs/1902.06890}
  {arXiv:1902.06890 [hep-ph]} \BibitemShut {NoStop}%
\bibitem [{\citenamefont {Kaplan}(1991)}]{Kaplan:1991dc}%
  \BibitemOpen
  \bibfield  {author} {\bibinfo {author} {\bibfnamefont {D.~B.}\ \bibnamefont
  {Kaplan}},\ }\href {\doibase 10.1016/S0550-3213(05)80021-5} {\bibfield
  {journal} {\bibinfo  {journal} {Nucl. Phys.}\ }\textbf {\bibinfo {volume}
  {B365}},\ \bibinfo {pages} {259} (\bibinfo {year} {1991})}\BibitemShut
  {NoStop}%
\bibitem [{\citenamefont {Golterman}\ and\ \citenamefont
  {Shamir}(2015)}]{Golterman:2015zwa}%
  \BibitemOpen
  \bibfield  {author} {\bibinfo {author} {\bibfnamefont {M.}~\bibnamefont
  {Golterman}}\ and\ \bibinfo {author} {\bibfnamefont {Y.}~\bibnamefont
  {Shamir}},\ }\href {\doibase 10.1103/PhysRevD.91.094506} {\bibfield
  {journal} {\bibinfo  {journal} {Phys. Rev.}\ }\textbf {\bibinfo {volume}
  {D91}},\ \bibinfo {pages} {094506} (\bibinfo {year} {2015})},\ \Eprint
  {http://arxiv.org/abs/1502.00390} {arXiv:1502.00390 [hep-ph]} \BibitemShut
  {NoStop}%
\bibitem [{\citenamefont {Golterman}\ and\ \citenamefont
  {Shamir}(2018)}]{Golterman:2017vdj}%
  \BibitemOpen
  \bibfield  {author} {\bibinfo {author} {\bibfnamefont {M.}~\bibnamefont
  {Golterman}}\ and\ \bibinfo {author} {\bibfnamefont {Y.}~\bibnamefont
  {Shamir}},\ }\href {\doibase 10.1103/PhysRevD.97.095005} {\bibfield
  {journal} {\bibinfo  {journal} {Phys. Rev.}\ }\textbf {\bibinfo {volume}
  {D97}},\ \bibinfo {pages} {095005} (\bibinfo {year} {2018})},\ \Eprint
  {http://arxiv.org/abs/1707.06033} {arXiv:1707.06033 [hep-ph]} \BibitemShut
  {NoStop}%
\bibitem [{\citenamefont {Golterman}\ and\ \citenamefont
  {Shamir}(2014)}]{Golterman:2014yha}%
  \BibitemOpen
  \bibfield  {author} {\bibinfo {author} {\bibfnamefont {M.}~\bibnamefont
  {Golterman}}\ and\ \bibinfo {author} {\bibfnamefont {Y.}~\bibnamefont
  {Shamir}},\ }\href {\doibase 10.1103/PhysRevD.89.054501} {\bibfield
  {journal} {\bibinfo  {journal} {Phys. Rev.}\ }\textbf {\bibinfo {volume}
  {D89}},\ \bibinfo {pages} {054501} (\bibinfo {year} {2014})},\ \Eprint
  {http://arxiv.org/abs/1401.0356} {arXiv:1401.0356 [hep-lat]} \BibitemShut
  {NoStop}%
\bibitem [{\citenamefont {Peskin}(1980)}]{Peskin:1980gc}%
  \BibitemOpen
  \bibfield  {author} {\bibinfo {author} {\bibfnamefont {M.~E.}\ \bibnamefont
  {Peskin}},\ }\href {\doibase 10.1016/0550-3213(80)90051-6} {\bibfield
  {journal} {\bibinfo  {journal} {Nucl. Phys.}\ }\textbf {\bibinfo {volume}
  {B175}},\ \bibinfo {pages} {197} (\bibinfo {year} {1980})}\BibitemShut
  {NoStop}%
\bibitem [{\citenamefont {Witten}(1983)}]{Witten:1983ut}%
  \BibitemOpen
  \bibfield  {author} {\bibinfo {author} {\bibfnamefont {E.}~\bibnamefont
  {Witten}},\ }\href {\doibase 10.1103/PhysRevLett.51.2351} {\bibfield
  {journal} {\bibinfo  {journal} {Phys. Rev. Lett.}\ }\textbf {\bibinfo
  {volume} {51}},\ \bibinfo {pages} {2351} (\bibinfo {year}
  {1983})}\BibitemShut {NoStop}%
\bibitem [{\citenamefont {Agashe}\ \emph {et~al.}(2005)\citenamefont {Agashe},
  \citenamefont {Contino},\ and\ \citenamefont {Pomarol}}]{Agashe:2004rs}%
  \BibitemOpen
  \bibfield  {author} {\bibinfo {author} {\bibfnamefont {K.}~\bibnamefont
  {Agashe}}, \bibinfo {author} {\bibfnamefont {R.}~\bibnamefont {Contino}}, \
  and\ \bibinfo {author} {\bibfnamefont {A.}~\bibnamefont {Pomarol}},\ }\href
  {\doibase 10.1016/j.nuclphysb.2005.04.035} {\bibfield  {journal} {\bibinfo
  {journal} {Nucl. Phys.}\ }\textbf {\bibinfo {volume} {B719}},\ \bibinfo
  {pages} {165} (\bibinfo {year} {2005})},\ \Eprint
  {http://arxiv.org/abs/hep-ph/0412089} {arXiv:hep-ph/0412089 [hep-ph]}
  \BibitemShut {NoStop}%
\bibitem [{\citenamefont {Del~Debbio}\ \emph {et~al.}(2017)\citenamefont
  {Del~Debbio}, \citenamefont {Englert},\ and\ \citenamefont
  {Zwicky}}]{DelDebbio:2017ini}%
  \BibitemOpen
  \bibfield  {author} {\bibinfo {author} {\bibfnamefont {L.}~\bibnamefont
  {Del~Debbio}}, \bibinfo {author} {\bibfnamefont {C.}~\bibnamefont {Englert}},
  \ and\ \bibinfo {author} {\bibfnamefont {R.}~\bibnamefont {Zwicky}},\ }\href
  {\doibase 10.1007/JHEP08(2017)142} {\bibfield  {journal} {\bibinfo  {journal}
  {JHEP}\ }\textbf {\bibinfo {volume} {08}},\ \bibinfo {pages} {142} (\bibinfo
  {year} {2017})},\ \Eprint {http://arxiv.org/abs/1703.06064} {arXiv:1703.06064
  [hep-ph]} \BibitemShut {NoStop}%
\bibitem [{\citenamefont {Das}\ \emph {et~al.}(1967)\citenamefont {Das},
  \citenamefont {Guralnik}, \citenamefont {Mathur}, \citenamefont {Low},\ and\
  \citenamefont {Young}}]{Das:1967it}%
  \BibitemOpen
  \bibfield  {author} {\bibinfo {author} {\bibfnamefont {T.}~\bibnamefont
  {Das}}, \bibinfo {author} {\bibfnamefont {G.~S.}\ \bibnamefont {Guralnik}},
  \bibinfo {author} {\bibfnamefont {V.~S.}\ \bibnamefont {Mathur}}, \bibinfo
  {author} {\bibfnamefont {F.~E.}\ \bibnamefont {Low}}, \ and\ \bibinfo
  {author} {\bibfnamefont {J.~E.}\ \bibnamefont {Young}},\ }\href {\doibase
  10.1103/PhysRevLett.18.759} {\bibfield  {journal} {\bibinfo  {journal} {Phys.
  Rev. Lett.}\ }\textbf {\bibinfo {volume} {18}},\ \bibinfo {pages} {759}
  (\bibinfo {year} {1967})}\BibitemShut {NoStop}%
\bibitem [{\citenamefont {DeGrand}\ \emph {et~al.}(2016)\citenamefont
  {DeGrand}, \citenamefont {Golterman}, \citenamefont {Jay}, \citenamefont
  {Neil}, \citenamefont {Shamir},\ and\ \citenamefont
  {Svetitsky}}]{DeGrand:2016htl}%
  \BibitemOpen
  \bibfield  {author} {\bibinfo {author} {\bibfnamefont {T.~A.}\ \bibnamefont
  {DeGrand}}, \bibinfo {author} {\bibfnamefont {M.}~\bibnamefont {Golterman}},
  \bibinfo {author} {\bibfnamefont {W.~I.}\ \bibnamefont {Jay}}, \bibinfo
  {author} {\bibfnamefont {E.~T.}\ \bibnamefont {Neil}}, \bibinfo {author}
  {\bibfnamefont {Y.}~\bibnamefont {Shamir}}, \ and\ \bibinfo {author}
  {\bibfnamefont {B.}~\bibnamefont {Svetitsky}},\ }\href {\doibase
  10.1103/PhysRevD.94.054501} {\bibfield  {journal} {\bibinfo  {journal} {Phys.
  Rev.}\ }\textbf {\bibinfo {volume} {D94}},\ \bibinfo {pages} {054501}
  (\bibinfo {year} {2016})},\ \Eprint {http://arxiv.org/abs/1606.02695}
  {arXiv:1606.02695 [hep-lat]} \BibitemShut {NoStop}%
\bibitem [{\citenamefont {Shintani}\ \emph {et~al.}(2008)\citenamefont
  {Shintani}, \citenamefont {Aoki}, \citenamefont {Fukaya}, \citenamefont
  {Hashimoto}, \citenamefont {Kaneko}, \citenamefont {Matsufuru}, \citenamefont
  {Onogi},\ and\ \citenamefont {Yamada}}]{Shintani:2008qe}%
  \BibitemOpen
  \bibfield  {author} {\bibinfo {author} {\bibfnamefont {E.}~\bibnamefont
  {Shintani}}, \bibinfo {author} {\bibfnamefont {S.}~\bibnamefont {Aoki}},
  \bibinfo {author} {\bibfnamefont {H.}~\bibnamefont {Fukaya}}, \bibinfo
  {author} {\bibfnamefont {S.}~\bibnamefont {Hashimoto}}, \bibinfo {author}
  {\bibfnamefont {T.}~\bibnamefont {Kaneko}}, \bibinfo {author} {\bibfnamefont
  {H.}~\bibnamefont {Matsufuru}}, \bibinfo {author} {\bibfnamefont
  {T.}~\bibnamefont {Onogi}}, \ and\ \bibinfo {author} {\bibfnamefont
  {N.}~\bibnamefont {Yamada}} (\bibinfo {collaboration} {JLQCD}),\ }\href
  {\doibase 10.1103/PhysRevLett.101.242001} {\bibfield  {journal} {\bibinfo
  {journal} {Phys. Rev. Lett.}\ }\textbf {\bibinfo {volume} {101}},\ \bibinfo
  {pages} {242001} (\bibinfo {year} {2008})},\ \Eprint
  {http://arxiv.org/abs/0806.4222} {arXiv:0806.4222 [hep-lat]} \BibitemShut
  {NoStop}%
\bibitem [{\citenamefont {Boyle}\ \emph {et~al.}(2010)\citenamefont {Boyle},
  \citenamefont {Del~Debbio}, \citenamefont {Wennekers},\ and\ \citenamefont
  {Zanotti}}]{Boyle:2009xi}%
  \BibitemOpen
  \bibfield  {author} {\bibinfo {author} {\bibfnamefont {P.~A.}\ \bibnamefont
  {Boyle}}, \bibinfo {author} {\bibfnamefont {L.}~\bibnamefont {Del~Debbio}},
  \bibinfo {author} {\bibfnamefont {J.}~\bibnamefont {Wennekers}}, \ and\
  \bibinfo {author} {\bibfnamefont {J.~M.}\ \bibnamefont {Zanotti}} (\bibinfo
  {collaboration} {RBC, UKQCD}),\ }\href {\doibase 10.1103/PhysRevD.81.014504}
  {\bibfield  {journal} {\bibinfo  {journal} {Phys. Rev.}\ }\textbf {\bibinfo
  {volume} {D81}},\ \bibinfo {pages} {014504} (\bibinfo {year} {2010})},\
  \Eprint {http://arxiv.org/abs/0909.4931} {arXiv:0909.4931 [hep-lat]}
  \BibitemShut {NoStop}%
\bibitem [{\citenamefont {Appelquist}\ \emph {et~al.}(2011)\citenamefont
  {Appelquist} \emph {et~al.}}]{Appelquist:2010xv}%
  \BibitemOpen
  \bibfield  {author} {\bibinfo {author} {\bibfnamefont {T.}~\bibnamefont
  {Appelquist}} \emph {et~al.} (\bibinfo {collaboration} {LSD}),\ }\href
  {\doibase 10.1103/PhysRevLett.106.231601} {\bibfield  {journal} {\bibinfo
  {journal} {Phys. Rev. Lett.}\ }\textbf {\bibinfo {volume} {106}},\ \bibinfo
  {pages} {231601} (\bibinfo {year} {2011})},\ \Eprint
  {http://arxiv.org/abs/1009.5967} {arXiv:1009.5967 [hep-ph]} \BibitemShut
  {NoStop}%
\bibitem [{\citenamefont {Aubin}\ and\ \citenamefont
  {Blum}(2007)}]{Aubin:2006xv}%
  \BibitemOpen
  \bibfield  {author} {\bibinfo {author} {\bibfnamefont {C.}~\bibnamefont
  {Aubin}}\ and\ \bibinfo {author} {\bibfnamefont {T.}~\bibnamefont {Blum}},\
  }\href {\doibase 10.1103/PhysRevD.75.114502} {\bibfield  {journal} {\bibinfo
  {journal} {Phys. Rev.}\ }\textbf {\bibinfo {volume} {D75}},\ \bibinfo {pages}
  {114502} (\bibinfo {year} {2007})},\ \Eprint
  {http://arxiv.org/abs/hep-lat/0608011} {arXiv:hep-lat/0608011 [hep-lat]}
  \BibitemShut {NoStop}%
\bibitem [{\citenamefont {Chakraborty}\ \emph {et~al.}(2014)\citenamefont
  {Chakraborty}, \citenamefont {Davies}, \citenamefont {Donald}, \citenamefont
  {Dowdall}, \citenamefont {Koponen}, \citenamefont {Lepage},\ and\
  \citenamefont {Teubner}}]{Chakraborty:2014mwa}%
  \BibitemOpen
  \bibfield  {author} {\bibinfo {author} {\bibfnamefont {B.}~\bibnamefont
  {Chakraborty}}, \bibinfo {author} {\bibfnamefont {C.~T.~H.}\ \bibnamefont
  {Davies}}, \bibinfo {author} {\bibfnamefont {G.~C.}\ \bibnamefont {Donald}},
  \bibinfo {author} {\bibfnamefont {R.~J.}\ \bibnamefont {Dowdall}}, \bibinfo
  {author} {\bibfnamefont {J.}~\bibnamefont {Koponen}}, \bibinfo {author}
  {\bibfnamefont {G.~P.}\ \bibnamefont {Lepage}}, \ and\ \bibinfo {author}
  {\bibfnamefont {T.}~\bibnamefont {Teubner}} (\bibinfo {collaboration}
  {HPQCD}),\ }\href {\doibase 10.1103/PhysRevD.89.114501} {\bibfield  {journal}
  {\bibinfo  {journal} {Phys. Rev.}\ }\textbf {\bibinfo {volume} {D89}},\
  \bibinfo {pages} {114501} (\bibinfo {year} {2014})},\ \Eprint
  {http://arxiv.org/abs/1403.1778} {arXiv:1403.1778 [hep-lat]} \BibitemShut
  {NoStop}%
\bibitem [{\citenamefont {Chakraborty}\ \emph {et~al.}(2017)\citenamefont
  {Chakraborty}, \citenamefont {Davies}, \citenamefont {de~Oliviera},
  \citenamefont {Koponen}, \citenamefont {Lepage},\ and\ \citenamefont {Van~de
  Water}}]{Chakraborty:2016mwy}%
  \BibitemOpen
  \bibfield  {author} {\bibinfo {author} {\bibfnamefont {B.}~\bibnamefont
  {Chakraborty}}, \bibinfo {author} {\bibfnamefont {C.~T.~H.}\ \bibnamefont
  {Davies}}, \bibinfo {author} {\bibfnamefont {P.~G.}\ \bibnamefont
  {de~Oliviera}}, \bibinfo {author} {\bibfnamefont {J.}~\bibnamefont
  {Koponen}}, \bibinfo {author} {\bibfnamefont {G.~P.}\ \bibnamefont {Lepage}},
  \ and\ \bibinfo {author} {\bibfnamefont {R.~S.}\ \bibnamefont {Van~de
  Water}},\ }\href {\doibase 10.1103/PhysRevD.96.034516} {\bibfield  {journal}
  {\bibinfo  {journal} {Phys. Rev.}\ }\textbf {\bibinfo {volume} {D96}},\
  \bibinfo {pages} {034516} (\bibinfo {year} {2017})},\ \Eprint
  {http://arxiv.org/abs/1601.03071} {arXiv:1601.03071 [hep-lat]} \BibitemShut
  {NoStop}%
\bibitem [{\citenamefont {Borsanyi}\ \emph {et~al.}(2018)\citenamefont
  {Borsanyi} \emph {et~al.}}]{Borsanyi:2017zdw}%
  \BibitemOpen
  \bibfield  {author} {\bibinfo {author} {\bibfnamefont {S.}~\bibnamefont
  {Borsanyi}} \emph {et~al.} (\bibinfo {collaboration}
  {Budapest-Marseille-Wuppertal}),\ }\href {\doibase
  10.1103/PhysRevLett.121.022002} {\bibfield  {journal} {\bibinfo  {journal}
  {Phys. Rev. Lett.}\ }\textbf {\bibinfo {volume} {121}},\ \bibinfo {pages}
  {022002} (\bibinfo {year} {2018})},\ \Eprint
  {http://arxiv.org/abs/1711.04980} {arXiv:1711.04980 [hep-lat]} \BibitemShut
  {NoStop}%
\bibitem [{\citenamefont {Aubin}\ \emph {et~al.}(2018)\citenamefont {Aubin},
  \citenamefont {Blum}, \citenamefont {Golterman}, \citenamefont {Jung},
  \citenamefont {Peris},\ and\ \citenamefont {Tu}}]{Aubin:2018fog}%
  \BibitemOpen
  \bibfield  {author} {\bibinfo {author} {\bibfnamefont {C.}~\bibnamefont
  {Aubin}}, \bibinfo {author} {\bibfnamefont {T.}~\bibnamefont {Blum}},
  \bibinfo {author} {\bibfnamefont {M.}~\bibnamefont {Golterman}}, \bibinfo
  {author} {\bibfnamefont {C.}~\bibnamefont {Jung}}, \bibinfo {author}
  {\bibfnamefont {S.}~\bibnamefont {Peris}}, \ and\ \bibinfo {author}
  {\bibfnamefont {C.}~\bibnamefont {Tu}},\ }in\ \href@noop {} {\emph {\bibinfo
  {booktitle} {{36th International Symposium on Lattice Field Theory (Lattice
  2018) East Lansing, MI, United States, July 22-28, 2018}}}}\ (\bibinfo {year}
  {2018})\ \Eprint {http://arxiv.org/abs/1812.03334} {arXiv:1812.03334
  [hep-lat]} \BibitemShut {NoStop}%
\bibitem [{\citenamefont {Ayyar}\ \emph
  {et~al.}(2018{\natexlab{a}})\citenamefont {Ayyar}, \citenamefont {DeGrand},
  \citenamefont {Golterman}, \citenamefont {Hackett}, \citenamefont {Jay},
  \citenamefont {Neil}, \citenamefont {Shamir},\ and\ \citenamefont
  {Svetitsky}}]{Ayyar:2017qdf}%
  \BibitemOpen
  \bibfield  {author} {\bibinfo {author} {\bibfnamefont {V.}~\bibnamefont
  {Ayyar}}, \bibinfo {author} {\bibfnamefont {T.}~\bibnamefont {DeGrand}},
  \bibinfo {author} {\bibfnamefont {M.}~\bibnamefont {Golterman}}, \bibinfo
  {author} {\bibfnamefont {D.~C.}\ \bibnamefont {Hackett}}, \bibinfo {author}
  {\bibfnamefont {W.~I.}\ \bibnamefont {Jay}}, \bibinfo {author} {\bibfnamefont
  {E.~T.}\ \bibnamefont {Neil}}, \bibinfo {author} {\bibfnamefont
  {Y.}~\bibnamefont {Shamir}}, \ and\ \bibinfo {author} {\bibfnamefont
  {B.}~\bibnamefont {Svetitsky}},\ }\href {\doibase 10.1103/PhysRevD.97.074505}
  {\bibfield  {journal} {\bibinfo  {journal} {Phys. Rev.}\ }\textbf {\bibinfo
  {volume} {D97}},\ \bibinfo {pages} {074505} (\bibinfo {year}
  {2018}{\natexlab{a}})},\ \Eprint {http://arxiv.org/abs/1710.00806}
  {arXiv:1710.00806 [hep-lat]} \BibitemShut {NoStop}%
\bibitem [{\citenamefont {Ayyar}\ \emph
  {et~al.}(2018{\natexlab{b}})\citenamefont {Ayyar}, \citenamefont {Degrand},
  \citenamefont {Hackett}, \citenamefont {Jay}, \citenamefont {Neil},
  \citenamefont {Shamir},\ and\ \citenamefont {Svetitsky}}]{Ayyar:2018zuk}%
  \BibitemOpen
  \bibfield  {author} {\bibinfo {author} {\bibfnamefont {V.}~\bibnamefont
  {Ayyar}}, \bibinfo {author} {\bibfnamefont {T.}~\bibnamefont {Degrand}},
  \bibinfo {author} {\bibfnamefont {D.~C.}\ \bibnamefont {Hackett}}, \bibinfo
  {author} {\bibfnamefont {W.~I.}\ \bibnamefont {Jay}}, \bibinfo {author}
  {\bibfnamefont {E.~T.}\ \bibnamefont {Neil}}, \bibinfo {author}
  {\bibfnamefont {Y.}~\bibnamefont {Shamir}}, \ and\ \bibinfo {author}
  {\bibfnamefont {B.}~\bibnamefont {Svetitsky}},\ }\href {\doibase
  10.1103/PhysRevD.97.114505} {\bibfield  {journal} {\bibinfo  {journal} {Phys.
  Rev.}\ }\textbf {\bibinfo {volume} {D97}},\ \bibinfo {pages} {114505}
  (\bibinfo {year} {2018}{\natexlab{b}})},\ \Eprint
  {http://arxiv.org/abs/1801.05809} {arXiv:1801.05809 [hep-ph]} \BibitemShut
  {NoStop}%
\bibitem [{\citenamefont {Ayyar}\ \emph
  {et~al.}(2018{\natexlab{c}})\citenamefont {Ayyar}, \citenamefont {DeGrand},
  \citenamefont {Hackett}, \citenamefont {Jay}, \citenamefont {Neil},
  \citenamefont {Shamir},\ and\ \citenamefont {Svetitsky}}]{Ayyar:2018glg}%
  \BibitemOpen
  \bibfield  {author} {\bibinfo {author} {\bibfnamefont {V.}~\bibnamefont
  {Ayyar}}, \bibinfo {author} {\bibfnamefont {T.}~\bibnamefont {DeGrand}},
  \bibinfo {author} {\bibfnamefont {D.~C.}\ \bibnamefont {Hackett}}, \bibinfo
  {author} {\bibfnamefont {W.~I.}\ \bibnamefont {Jay}}, \bibinfo {author}
  {\bibfnamefont {E.~T.}\ \bibnamefont {Neil}}, \bibinfo {author}
  {\bibfnamefont {Y.}~\bibnamefont {Shamir}}, \ and\ \bibinfo {author}
  {\bibfnamefont {B.}~\bibnamefont {Svetitsky}},\ }\href@noop {} {\  (\bibinfo
  {year} {2018}{\natexlab{c}})},\ \Eprint {http://arxiv.org/abs/1812.02727}
  {arXiv:1812.02727 [hep-ph]} \BibitemShut {NoStop}%
\bibitem [{\citenamefont {DeGrand}\ and\ \citenamefont
  {DeTar}(2006)}]{DeGrand:2006zz}%
  \BibitemOpen
  \bibfield  {author} {\bibinfo {author} {\bibfnamefont {T.}~\bibnamefont
  {DeGrand}}\ and\ \bibinfo {author} {\bibfnamefont {C.~E.}\ \bibnamefont
  {DeTar}},\ }\href@noop {} {\emph {\bibinfo {title} {{Lattice methods for
  quantum chromodynamics}}}}\ (\bibinfo {year} {2006})\BibitemShut {NoStop}%
\bibitem [{\citenamefont {Hasenfratz}\ and\ \citenamefont
  {Knechtli}(2001)}]{Hasenfratz:2001hp}%
  \BibitemOpen
  \bibfield  {author} {\bibinfo {author} {\bibfnamefont {A.}~\bibnamefont
  {Hasenfratz}}\ and\ \bibinfo {author} {\bibfnamefont {F.}~\bibnamefont
  {Knechtli}},\ }\href {\doibase 10.1103/PhysRevD.64.034504} {\bibfield
  {journal} {\bibinfo  {journal} {Phys. Rev.}\ }\textbf {\bibinfo {volume}
  {D64}},\ \bibinfo {pages} {034504} (\bibinfo {year} {2001})},\ \Eprint
  {http://arxiv.org/abs/hep-lat/0103029} {arXiv:hep-lat/0103029 [hep-lat]}
  \BibitemShut {NoStop}%
\bibitem [{\citenamefont {Hasenfratz}\ \emph {et~al.}(2007)\citenamefont
  {Hasenfratz}, \citenamefont {Hoffmann},\ and\ \citenamefont
  {Schaefer}}]{Hasenfratz:2007rf}%
  \BibitemOpen
  \bibfield  {author} {\bibinfo {author} {\bibfnamefont {A.}~\bibnamefont
  {Hasenfratz}}, \bibinfo {author} {\bibfnamefont {R.}~\bibnamefont
  {Hoffmann}}, \ and\ \bibinfo {author} {\bibfnamefont {S.}~\bibnamefont
  {Schaefer}},\ }\href {\doibase 10.1088/1126-6708/2007/05/029} {\bibfield
  {journal} {\bibinfo  {journal} {JHEP}\ }\textbf {\bibinfo {volume} {05}},\
  \bibinfo {pages} {029} (\bibinfo {year} {2007})},\ \Eprint
  {http://arxiv.org/abs/hep-lat/0702028} {arXiv:hep-lat/0702028 [hep-lat]}
  \BibitemShut {NoStop}%
\bibitem [{\citenamefont {Bernard}\ and\ \citenamefont
  {DeGrand}(2000)}]{Bernard:1999kc}%
  \BibitemOpen
  \bibfield  {author} {\bibinfo {author} {\bibfnamefont {C.~W.}\ \bibnamefont
  {Bernard}}\ and\ \bibinfo {author} {\bibfnamefont {T.~A.}\ \bibnamefont
  {DeGrand}},\ }\bibfield  {booktitle} {\emph {\bibinfo {booktitle} {{Lattice
  field theory. Proceedings, 17th International Symposium, Lattice'99, Pisa,
  Italy, June 29-July 3, 1999}}},\ }\href {\doibase
  10.1016/S0920-5632(00)91822-X} {\bibfield  {journal} {\bibinfo  {journal}
  {Nucl. Phys. Proc. Suppl.}\ }\textbf {\bibinfo {volume} {83}},\ \bibinfo
  {pages} {845} (\bibinfo {year} {2000})},\ \Eprint
  {http://arxiv.org/abs/hep-lat/9909083} {arXiv:hep-lat/9909083 [hep-lat]}
  \BibitemShut {NoStop}%
\bibitem [{\citenamefont {Shamir}\ \emph {et~al.}(2011)\citenamefont {Shamir},
  \citenamefont {Svetitsky},\ and\ \citenamefont {Yurkovsky}}]{Shamir:2010cq}%
  \BibitemOpen
  \bibfield  {author} {\bibinfo {author} {\bibfnamefont {Y.}~\bibnamefont
  {Shamir}}, \bibinfo {author} {\bibfnamefont {B.}~\bibnamefont {Svetitsky}}, \
  and\ \bibinfo {author} {\bibfnamefont {E.}~\bibnamefont {Yurkovsky}},\ }\href
  {\doibase 10.1103/PhysRevD.83.097502} {\bibfield  {journal} {\bibinfo
  {journal} {Phys. Rev.}\ }\textbf {\bibinfo {volume} {D83}},\ \bibinfo {pages}
  {097502} (\bibinfo {year} {2011})},\ \Eprint {http://arxiv.org/abs/1012.2819}
  {arXiv:1012.2819 [hep-lat]} \BibitemShut {NoStop}%
\bibitem [{\citenamefont {DeGrand}\ \emph {et~al.}(2014)\citenamefont
  {DeGrand}, \citenamefont {Shamir},\ and\ \citenamefont
  {Svetitsky}}]{DeGrand:2014rwa}%
  \BibitemOpen
  \bibfield  {author} {\bibinfo {author} {\bibfnamefont {T.}~\bibnamefont
  {DeGrand}}, \bibinfo {author} {\bibfnamefont {Y.}~\bibnamefont {Shamir}}, \
  and\ \bibinfo {author} {\bibfnamefont {B.}~\bibnamefont {Svetitsky}},\ }\href
  {\doibase 10.1103/PhysRevD.90.054501} {\bibfield  {journal} {\bibinfo
  {journal} {Phys. Rev.}\ }\textbf {\bibinfo {volume} {D90}},\ \bibinfo {pages}
  {054501} (\bibinfo {year} {2014})},\ \Eprint {http://arxiv.org/abs/1407.4201}
  {arXiv:1407.4201 [hep-lat]} \BibitemShut {NoStop}%
\bibitem [{\citenamefont {Golterman}(1986)}]{Golterman:1985dz}%
  \BibitemOpen
  \bibfield  {author} {\bibinfo {author} {\bibfnamefont {M.~F.~L.}\
  \bibnamefont {Golterman}},\ }\href {\doibase 10.1016/0550-3213(86)90383-4}
  {\bibfield  {journal} {\bibinfo  {journal} {Nucl. Phys.}\ }\textbf {\bibinfo
  {volume} {B273}},\ \bibinfo {pages} {663} (\bibinfo {year}
  {1986})}\BibitemShut {NoStop}%
\bibitem [{\citenamefont {Lee}\ and\ \citenamefont
  {Sharpe}(1999)}]{Lee:1999zxa}%
  \BibitemOpen
  \bibfield  {author} {\bibinfo {author} {\bibfnamefont {W.-J.}\ \bibnamefont
  {Lee}}\ and\ \bibinfo {author} {\bibfnamefont {S.~R.}\ \bibnamefont
  {Sharpe}},\ }\href {\doibase 10.1103/PhysRevD.60.114503} {\bibfield
  {journal} {\bibinfo  {journal} {Phys. Rev.}\ }\textbf {\bibinfo {volume}
  {D60}},\ \bibinfo {pages} {114503} (\bibinfo {year} {1999})},\ \Eprint
  {http://arxiv.org/abs/hep-lat/9905023} {arXiv:hep-lat/9905023 [hep-lat]}
  \BibitemShut {NoStop}%
\bibitem [{\citenamefont {Aubin}\ \emph {et~al.}(2016)\citenamefont {Aubin},
  \citenamefont {Blum}, \citenamefont {Chau}, \citenamefont {Golterman},
  \citenamefont {Peris},\ and\ \citenamefont {Tu}}]{Aubin:2015rzx}%
  \BibitemOpen
  \bibfield  {author} {\bibinfo {author} {\bibfnamefont {C.}~\bibnamefont
  {Aubin}}, \bibinfo {author} {\bibfnamefont {T.}~\bibnamefont {Blum}},
  \bibinfo {author} {\bibfnamefont {P.}~\bibnamefont {Chau}}, \bibinfo {author}
  {\bibfnamefont {M.}~\bibnamefont {Golterman}}, \bibinfo {author}
  {\bibfnamefont {S.}~\bibnamefont {Peris}}, \ and\ \bibinfo {author}
  {\bibfnamefont {C.}~\bibnamefont {Tu}},\ }\href {\doibase
  10.1103/PhysRevD.93.054508} {\bibfield  {journal} {\bibinfo  {journal} {Phys.
  Rev.}\ }\textbf {\bibinfo {volume} {D93}},\ \bibinfo {pages} {054508}
  (\bibinfo {year} {2016})},\ \Eprint {http://arxiv.org/abs/1512.07555}
  {arXiv:1512.07555 [hep-lat]} \BibitemShut {NoStop}%
\bibitem [{\citenamefont {B{\"a}r}\ \emph {et~al.}(2011)\citenamefont
  {B{\"a}r}, \citenamefont {Golterman},\ and\ \citenamefont
  {Shamir}}]{Bar:2010ix}%
  \BibitemOpen
  \bibfield  {author} {\bibinfo {author} {\bibfnamefont {O.}~\bibnamefont
  {B{\"a}r}}, \bibinfo {author} {\bibfnamefont {M.}~\bibnamefont {Golterman}},
  \ and\ \bibinfo {author} {\bibfnamefont {Y.}~\bibnamefont {Shamir}},\ }\href
  {\doibase 10.1103/PhysRevD.83.054501} {\bibfield  {journal} {\bibinfo
  {journal} {Phys. Rev.}\ }\textbf {\bibinfo {volume} {D83}},\ \bibinfo {pages}
  {054501} (\bibinfo {year} {2011})},\ \Eprint {http://arxiv.org/abs/1012.0987}
  {arXiv:1012.0987 [hep-lat]} \BibitemShut {NoStop}%
\bibitem [{\citenamefont {Bennett}\ \emph {et~al.}(2018)\citenamefont
  {Bennett}, \citenamefont {Hong}, \citenamefont {Lee}, \citenamefont {Lin},
  \citenamefont {Lucini}, \citenamefont {Piai},\ and\ \citenamefont
  {Vadacchino}}]{Bennett:2017kga}%
  \BibitemOpen
  \bibfield  {author} {\bibinfo {author} {\bibfnamefont {E.}~\bibnamefont
  {Bennett}}, \bibinfo {author} {\bibfnamefont {D.~K.}\ \bibnamefont {Hong}},
  \bibinfo {author} {\bibfnamefont {J.-W.}\ \bibnamefont {Lee}}, \bibinfo
  {author} {\bibfnamefont {C.~J.~D.}\ \bibnamefont {Lin}}, \bibinfo {author}
  {\bibfnamefont {B.}~\bibnamefont {Lucini}}, \bibinfo {author} {\bibfnamefont
  {M.}~\bibnamefont {Piai}}, \ and\ \bibinfo {author} {\bibfnamefont
  {D.}~\bibnamefont {Vadacchino}},\ }\href {\doibase 10.1007/JHEP03(2018)185}
  {\bibfield  {journal} {\bibinfo  {journal} {JHEP}\ }\textbf {\bibinfo
  {volume} {03}},\ \bibinfo {pages} {185} (\bibinfo {year} {2018})},\ \Eprint
  {http://arxiv.org/abs/1712.04220} {arXiv:1712.04220 [hep-lat]} \BibitemShut
  {NoStop}%
\bibitem [{\citenamefont {Lee}\ \emph {et~al.}(2018)\citenamefont {Lee},
  \citenamefont {Bennett}, \citenamefont {Hong}, \citenamefont {Lin},
  \citenamefont {Lucini}, \citenamefont {Piai},\ and\ \citenamefont
  {Vadacchino}}]{Lee:2018ztv}%
  \BibitemOpen
  \bibfield  {author} {\bibinfo {author} {\bibfnamefont {J.-W.}\ \bibnamefont
  {Lee}}, \bibinfo {author} {\bibnamefont {Bennett}}, \bibinfo {author}
  {\bibfnamefont {D.~K.}\ \bibnamefont {Hong}}, \bibinfo {author}
  {\bibfnamefont {C.~J.~D.}\ \bibnamefont {Lin}}, \bibinfo {author}
  {\bibfnamefont {B.}~\bibnamefont {Lucini}}, \bibinfo {author} {\bibfnamefont
  {M.}~\bibnamefont {Piai}}, \ and\ \bibinfo {author} {\bibfnamefont
  {D.}~\bibnamefont {Vadacchino}},\ }\bibfield  {booktitle} {\emph {\bibinfo
  {booktitle} {{36th International Symposium on Lattice Field Theory (Lattice
  2018) East Lansing, MI, United States, July 22-28, 2018}}},\ }\href@noop {}
  {\bibfield  {journal} {\bibinfo  {journal} {PoS}\ }\textbf {\bibinfo {volume}
  {LATTICE2018}},\ \bibinfo {pages} {192} (\bibinfo {year} {2018})},\ \Eprint
  {http://arxiv.org/abs/1811.00276} {arXiv:1811.00276 [hep-lat]} \BibitemShut
  {NoStop}%
\bibitem [{\citenamefont {{MILC Collaboration}}()}]{MILC}%
  \BibitemOpen
  \bibfield  {author} {\bibinfo {author} {\bibnamefont {{MILC
  Collaboration}}},\ }\href@noop {} {}\bibinfo {howpublished}
  {\url{http://www.physics.utah.edu/~detar/milc/}}\BibitemShut {NoStop}%
\end{thebibliography}%

\end{document}